\documentclass[prb,twocolumn,superscriptaddress,floatfix]{revtex4}
\usepackage{graphicx}
\usepackage{amssymb}
\usepackage{amsmath}
\usepackage{xspace} 
\newcommand{\e}{\mathrm{e}}
\newcommand{\D}{\mathrm{d}}
\newcommand{\imag}{\mathrm{i}}
\begin{document}
\title{Spatiotemporal evolution of polaronic states in finite quantum systems}
\author{H.~Fehske}
\affiliation{
Institut f\"ur Physik, Ernst-Moritz-Arndt-Universit{\"a}t
Greifswald, 17487 Greifswald, Germany }
\author{G. Wellein}
\affiliation{Regionales Rechenzentrum Erlangen, 
Universit\"at Erlangen-N\"urnberg, 91058 Erlangen, Germany }
\author{A. R. Bishop}
{\affiliation{\mbox{Theory, Simulation and Computation Directorate, 
Los Alamos National Laboratory, Los Alamos, New Mexico 87545}}
\begin{abstract}
We study the quantum dynamics of small polaron formation and polaron transport through finite  
quantum structures in the framework of the one-dimensional Holstein model 
with site-dependent potentials and interactions. Combining Lanczos diagonalization  
with Chebyshev moment expansion of the time evolution operator,
we determine how different initial states, representing stationary ground states or injected 
wave packets, after an electron-phonon interaction quench, develop in real space and time.  
Thereby, the full quantum nature and dynamics of electrons and phonons is preserved.
We find that the decay out of the initial state sensitively depends on the energy and momentum 
of the incoming particle, the electron-phonon coupling strength, and the phonon frequency,
whereupon bound polaron-phonon excited states may emerge in the strong-coupling regime.  
The tunneling of a Holstein polaron through a quantum wall or dot is generally accompanied 
by strong phonon number fluctuations due to phonon emission and reabsorption processes.
\end{abstract}
\pacs{73.63.-b,72.10.-d,71.38.-k,71.10.Fd}
\maketitle
\section{Introduction}
Electrons injected into low-dimensional quantum structures with strong electron-phonon (EP) 
interaction can cause local lattice deformations and thereby relax to ``self-trapped'' polaron states. 
Polaron self-trapping does not imply a breaking of translational invariance. When the polaron forms, 
the electron is being dressed by a phonon cloud, and the collective state
translates through the lattice. 
This indicates that vibrational  modes for polaron transport through 
nanoscale quantum devices are of vital importance.  The microscopic structure of polarons and the contexts in which they appear are rather diverse.
Phonon and polaron effects have been  investigated for , e.g., 
molecular transistors,~\cite{MAM04}
quantum dots,~\cite{IS92}
tunneling diodes and Aharonov-Bohm rings,~\cite{BT95}  
metal/organic/metal structures,~\cite{YSSB99} and Carbon 
nanotubes,~\cite{LLKD04,SBLH07}
although primarily with respect to steady state properties.
Recently, time-resolved spectroscopy has made it possible to address also the dynamical aspects 
of self-trapping, e.g., by directly time resolving the vibrational motions associated with the localized carrier.  
Taking advantage of the ultra-short pulse-widths of recent lasers, the femto-second dynamics of polaron 
formation and exciton-phonon dressing has been observed in pump-probe 
experiments.~\cite{TNSSTK98}

From a theoretical point of view, describing the time dependence  of small polaron formation 
requires a physics that is related to particle and phonon dynamics on the scale of the unit cell.~\cite{Ra06}
The simplest model that captures such a situation is the lattice-polaron Holstein model.~\cite{Ho59a} This model assumes that
the orbital states are identical on each site and the particle can move from site to site exactly as in a tight-binding model.
The phonons are coupled to the particle at whichever site it is on. The dynamics of the phonons is treated purely
locally with Einstein oscillators representing the intra-site (molecular) vibrations. After six decades of intense research
the equilibrium properties of the Holstein model are well-understood, at least in the single-particle sector     
(for recent reviews, see Refs.~\onlinecite{FT07,AD10}). In contrast, 
there is a rather incomplete understanding 
of time-dependent and out-of-equilibrium phenomena.  Some issues seem to be settled, e.g.,
the charge-transfer and correlated charge-deformation dynamics (but only for a two-site Holstein model),~\cite{MR97} the 
increase of the polaron formation time in two and three dimensions due to an adiabatic potential barrier
between extended electron and self-trapped polaron states,~\cite{MS77,KM93} or the conditional hopping 
rate of an injected electron and the vibrational relaxation time.~\cite{EK86} 
Quite recently the problem of determining the polaron formation time has 
been tackled.~\cite{KT07} Many questions remain at least partly unsolved 
however. For instance, in what way is quantum-dot polaron formation different in the adiabatic and anti-adiabatic 
regimes? And how does a bare particle evolves into a polaron after 
an ``interaction quench''? Or, how does a polaronic 
quasiparticle tunnel through a  potential barrier?   

In this paper,  we address some of these questions. To that end, we calculate---by means of 
numerically exact Lanczos diagonalization and Chebyshev expansion techniques---the real space and time 
evolution of polaronic states in the one-dimensional Holstein model 
with spatially and temporally varying on-site 
potentials and/or EP interaction strengths. The proposed approach 
is applicable to the dynamics of quasiparticle 
formation in several branches of physics.  

\section{Model and method}
\subsection{Modified Holstein Hamiltonian}
Focusing on polaron formation in one-dimensional finite quantum structures  
with short-range non-polar EP interaction we consider the
generalized Holstein molecular crystal model~\cite{Ho59a,FWLB08}
\begin{eqnarray}
\label{eq:ghm}
H &=& \sum_{i}\Delta_i n_{i} - t_0\sum_{i}
(c^{\dagger}_{i}c^{}_{i+1}
+\mbox{H.c.})\nonumber\\
&&-\sum_{i} g_i\omega_0
(b_i^{\dagger}+b_i^{})n_{i}+\omega_0\sum_{i} b_i^{\dagger} b_i^{}\,,
\end{eqnarray}
where $c_i^{\dagger}$ ($c_i^{}$) and $b_i^{\dagger}$ ($b_i^{}$) are creation (annihilation) operators for electrons and
dispersionless optical  phonons on site $i$, 
respectively, and 
$n_i^{}=c_i^{\dagger}c_i^{}$ is the corresponding particle number operator. 
In \eqref{eq:ghm}, the site-dependent potentials $\Delta_i$ 
can describe a tunnel barrier, a voltage bias, or disorder effects. 
$t_0$ denotes the nearest-neighbor electron transfer integral, 
and $g_i$ gives the local interaction of an electron on Wannier 
site $i$ to  an internal vibrational mode with frequency  $\omega_0$. 

The ratio $\omega_0/t_0$ determines which of the two subsystems, 
electrons or phonons, is the fast or the slow one. In the adiabatic 
limit $ (\omega_0/t_0)\ll 1$, 
the motion of the particle is affected by quasi-static lattice
deformations, whereas in the opposite, anti-adiabatic 
limit $(\omega_0/t_0)\gg 1$ the lattice deformation is presumed to
adjust instantaneously to the position of the carrier.

The dimensionless EP coupling constant $g^2$ normally appears 
in (small polaron) strong-coupling perturbation theory, where it 
describes the polaronic mass enhancement $m^*/m=e^{g^2}$
(for homogeneous systems, $g_i=g$). 
There is another natural measure of the strength of the 
electron-phonon interaction, the familiar polaronic level shift 
$E_p$. At strong EP coupling, $E_p$ gives the leading-order 
energy shift of the band dispersion.~\cite{AK99}  
In general, there is no simple relation between $g^2$
and $E_p$.  If the EP coupling is local and the phonon mode
is dispersionless, however, then $g^2=E_p/\omega_0$, and  
$E_p$ is usually identified with the polaron binding energy.~\cite{Zo99} 

The crossover from essentially free electronic carriers to 
heavy polaronic quasiparticles    
is known to occur for a translational invariant system, 
provided that two conditions, $g^2 > 1$ and $ E_{p}/zt_0 > 1$ 
($z=2$ ( in one dimension), 
are fulfilled.~\cite{CSG97}
So while the first requirement is more restrictive in the anti-adiabatic case, 
the formation of a small polaron state will be determined by the second 
criterion in the adiabatic regime. This likewise holds for the 
generalized Holstein model \eqref{eq:ghm} where  
$g_i^2=E_{p,i}/\omega_0$ [with a view to the different cases 
studied in Sec.~IIII we split 
$E_{p,i}=\varepsilon_{p}+\varepsilon_{p,i}$ up into
a constant ($\varepsilon_p$) and a site-dependent part
($\varepsilon_{p,i}$)].~\cite{FWLB08}

When investigating the physically most interesting crossover regime of the Holstein model where polarons form,
i.e. the self-trapping transition of the charge carriers takes place, standard analytical approaches 
fail to a large extent. This is because, precisely in this situation, the characteristic electronic and 
phononic energy scales are not well separated. So far quasi approximation-free numerical methods like 
quantum Monte Carlo simulations,~\cite{RL83}
exact diagonalizations,~\cite{Ma93}
or density-matrix renormalisation group 
techniques~\cite{JW98b}
yield the most reliable results for the ground-state 
and spectral  properties of Holstein polarons.

\subsection{Chebyshev expansion technique}
To study the real space and time formation of a polaronic quasiparticle from a bare electron
the time-dependent many-body Schr\"odinger equation has to be solved. For systems with moderate
Hilbert space  dimensions a full diagonalization of the Hamiltonian allows for an exact calculation 
of the quantum state at arbitrary times. Because of the phonon degrees of freedom the 
Hilbert space of the Holstein model is infinite, even for a finite lattice and in the single-particle sector.  
Truncating the Hilbert space of the 
phonons or constructing a variational Hilbert space including multiple-phonon excitations,~\cite{FT07,JF07}
a direct numerical integration of the Schr\"odinger equation can be performed, yielding
the  polaron many-body wave function at early times.~\cite{KT07} 
Alternatively one can exploit a
Chebyshev moment based expansion of the time evolution operator.~\cite{WF08} 
Since this technique also applies to very general situations and has been proven 
to be superior to direct integration and other iterative Schr\"odinger-equation solution schemes as to its efficiency 
(i.e. computational costs) and accuracy,~\cite{FSSWFB09}  the remainder of this section briefly outlines 
this less well-known approach.
     
The time evolution of a quantum state $|\psi\rangle$ is 
described by the Schr{\"o}dinger equation
\begin{equation}
  \imag \hbar \frac{\partial}{\partial t}|\psi(t)\rangle = H |\psi (t)\rangle\,.
\end{equation}
If the Hamilton operator $H$ does not explicitly depend on time $t$ 
we can formally integrate this equation and express 
the dynamics of an initial state $|\psi(0)\rangle$
in terms of the time evolution operator $U(t,0)$ as 
\begin{equation}
|\psi(t)\rangle = U(t,0)|\psi(0)\rangle\,,
\end{equation}
where 
\begin{equation}
U(t,0) = \e^{ -\imag H t/\hbar}\,.
\end{equation}

The time evolution operator $U(t+\Delta t,t) =U(\Delta t)$ for a given (usually small) time step $\Delta t$ 
can be expanded in a finite series of $N_{\rm C}$ first-kind Chebyshev polynomials of order $n$,  
\begin{equation}
T_n(x)=\cos [n\, \mathrm{arccos}(x)]\,.
\label{cheby}
\end{equation}
We obtain~\cite{TK84,WF08,CG99,FSSWFB09}
\begin{equation}
  U(\Delta t) =  \e^{-\imag b \Delta t/\hbar} 
  \Big[ c_0(a\Delta t/\hbar) + 2\sum\limits_{n=1}^{N_{\rm C}} c_n(a\Delta t/\hbar)
    T_n(\tilde{H}) \Big].
\label{eq:U_1}
\end{equation}

Prior to the expansion, the Hamiltonian has to be shifted and rescaled such that
the spectrum of $\tilde{H} = (H-b)/a$ is within the definition interval of the
Chebyshev polynomials, $[-1,1]$.~\cite{WWAF06}
The parameters $a$ and $b$ are calculated from the extremal eigenvalues of $H$ as
$b=\frac{1}{2}(E_{\mathrm{max}}+E_{\mathrm{min}})$ and 
$a=\frac{1}{2}(E_{\mathrm{max}}-E_{\mathrm{min}}+\epsilon)$.
Here we introduced $\epsilon=\alpha(E_{\mathrm{max}}-E_{\mathrm{min}})$ to ensure 
the rescaled spectrum $|\tilde{E}| \le 1/(1+\alpha)$ lies 
well inside $[-1,1]$. In practice, we use $\alpha=0.01$. 
The Chebyshev expansion also applies 
to systems with Holstein-type unbounded spectra.~\cite{WWAF06}
Here we can truncate the infinite Hilbert space to a finite 
dimension by restricting the model on a discrete space grid or using an energy
cutoff.
In this way we ensure the finiteness of the extreme eigenvalues.

In~(\ref{eq:U_1}), the expansion coefficients $c_n$ are given by 
\begin{equation}
  c_n(a\Delta t/\hbar) = \int\limits_{-1}^1 
  \frac{T_n(x)\e^{-\imag x a \Delta t/\hbar }}{\pi \sqrt{1-x^2}}\D x =
  (-\imag)^n J_n(a \Delta t/\hbar)\,;
\end{equation}
$J_n$ denotes the $n$-th order Bessel function of the first kind. 

In order to calculate the evolution of a state $|\psi(t)\rangle$ from one 
time grid point to the adjacent one,
\begin{equation}
|\psi(t+\Delta t)\rangle = U(\Delta t)|\psi(t)\rangle\,,
\end{equation}
we have to accumulate the $c_n$-weighted vectors
\begin{equation}
|v_n\rangle = T_n(\tilde{H})|\psi(t)\rangle\,.
\end{equation}
Since the coefficients $c_n(a\Delta t/\hbar)$ depend on the time step but not
on time explicitly, we need to calculate them only once.
Instead of evaluating Eq.~(\ref{cheby}) with $x=\tilde{H}$,
the vectors $|v_n\rangle$ can be computed iteratively exploiting
the recurrence relation of the Chebyshev polynomials,
\begin{equation}
    |v_{n+1}\rangle = 2\tilde{H} |v_n\rangle - |v_{n-1}\rangle\;,
\end{equation}
with $|v_1\rangle = \tilde{H} |v_0\rangle$ and $|v_0\rangle = |\psi(t)\rangle$.
Evolving the wave function from one time step to the next
requires $N_{\rm C}$  matrix vector multiplications  (MVMs) of a given 
complex vector with the sparse Hamilton matrix of dimension $D$ . 
Of course, to proceed from $t=0$ to $t$,
the procedure has to be performed  $t/\Delta t$ times.

Note that such a Chebyshev expansion may also be applied to systems with
time-dependent Hamiltonians, but there the time variation of 
$H(t)$ determines the maximum $\Delta t$ by which
the system may be propagated in a single time step.
For time-independent $H$, in principle, arbitrary large time steps are
possible at the expense of increasing $N_{\rm C}$.
We may choose $N_{\rm C}$ such that for $n>N_{\rm C}$ the modulus of all expansion coefficients 
\begin{equation}
|c_n(a\Delta t/\hbar)|\sim J_n(a\Delta t/\hbar)
\end{equation}
is smaller than a desired
accuracy cutoff.
This is facilitated by the fast asymptotic decay of the Bessel functions, 
\begin{equation}
  J_n(a\Delta t/\hbar)
  \sim \frac{1}{\sqrt{2\pi n}} \left( \frac{\e a\Delta t}{2\hbar n} \right)^n 
  \quad {\mathrm {for}}\qquad n\to \infty\;.
  \label{eq:dec_Bessel}
\end{equation}
Hence for $2\hbar n\gg \e a\Delta t$ the expansion coefficients $c_n$
decay superexponential and the series can be truncated with negligible 
error.~\cite{WF08} In the numerics of Sec.~III, we work with 
$N_{\rm C}\geq 10$, such that the last moment retained 
$|J_{N_{\rm C}}| \leq 10^{-9}$, i.e. the Chebyshev expansion 
can be considered as 
quasi-exact, and permits a considerably larger time step than
e.g. the Crank-Nicholson scheme.~\cite{PFTV86,FSSWFB09}
Of course, the ground-state energy $E_0(t)$ is unaltered 
during the simulation time.

Besides the high accuracy of the method, the linear scaling of 
computation time with both time step and Hilbert space dimension are 
promising in view of potential applications to more complex systems.
Here almost all computation time is spent in sparse MVMs, 
which can be efficiently parallelized, allowing for a good speedup on 
parallel computers. We use a memory saving implementation of the
MVM where the non-zero matrix elements
are not stored but recomputed in each sparse MVM step, limiting the
overall memory consumption of our implementation to five vectors of
size $D$. In this context we can access a massively parallel
sparse MVM code which has proven to be sufficient to compute the
ground state of the model~(\ref{eq:ghm}) up 
to $D=3.5\times10^{11}$ very efficiently on more than 5000
processor cores.~\cite{FAW09} For the single polaron dynamics presented
here, the matrix dimension is about $D=6\times10^8$ and we run the
Chebyshev approach on 18 processors of an SGI Altix4700 compute server,
accessing a total of approximately 60 GB of main memory and
consuming less than 1500 CPU-hrs to compute e.g. the results presented in
Sec.~III. C.

\section{Numerical Results and Discussion}
In this section we combine  exact diagonalization and Chebyshev expansion 
methods,~\cite{WWAF06,JF07,WF08}
working in the tensorial product Hilbert space of electrons and phonons.
We set $\hbar =1$ and give all energies in units of $t_0$. The time $t$ will be measured  with respect to 
the characteristic electronic and phononic  time scales $\tau_e=t_0^{-1}$ and 
$\tau_{ph}=(\omega_0/2\pi)^{-1}$, respectively.
We consider the case of a single electron only.
\subsection{Interaction quench}
To understand the basic features of the polaron formation process in the time domain, we first study a single 
oscillatory site to which the Holstein molecular crystal model applies, sandwiched between two ``wires'' 
where electrons are not coupled to phonons ($\epsilon_p=0$). 
The system size is $N=17$ with open boundary conditions 
(OBCs) at sites $i=1$, 17. We consider the case $\Delta_i=0$.
The deformable site is located midway, $i=9$.   Before time $t=0$ the system  is assumed to be in the non-interacting (free-electron) ground state;  its energy is $E_0=-1.9696$. Then, at $t=0$ the EP interaction at site $9$ is abruptly switched to a positive value $\varepsilon_{p,9}$ (an interaction ``quench''), 
what means that the electron and phonon subsystems are locally linked hereafter. 
Since the whole system is isolated from the environment  the total energy is conserved during the quench.

The time evolution of various quantities after such an interaction quench is shown in Figs.~\ref{fig:n_b_w0.3_9_wc}
to~\ref{fig:pdf_w80} for characteristic situations, ranging from weak to strong EP coupling  and 
adiabatic to anti-adiabatic cases. As can be seen from Figs.~\ref{fig:n_b_w0.3_9_wc}
to~\ref{fig:pdf_w80} the quantum dynamics after the quench depends on the EP coupling strength
and phonon frequency in a very sensitive way.

\subsubsection{Adiabatic regime} 

\begin{figure}[htbp]
\begin{center}
\includegraphics[width=.9\linewidth]{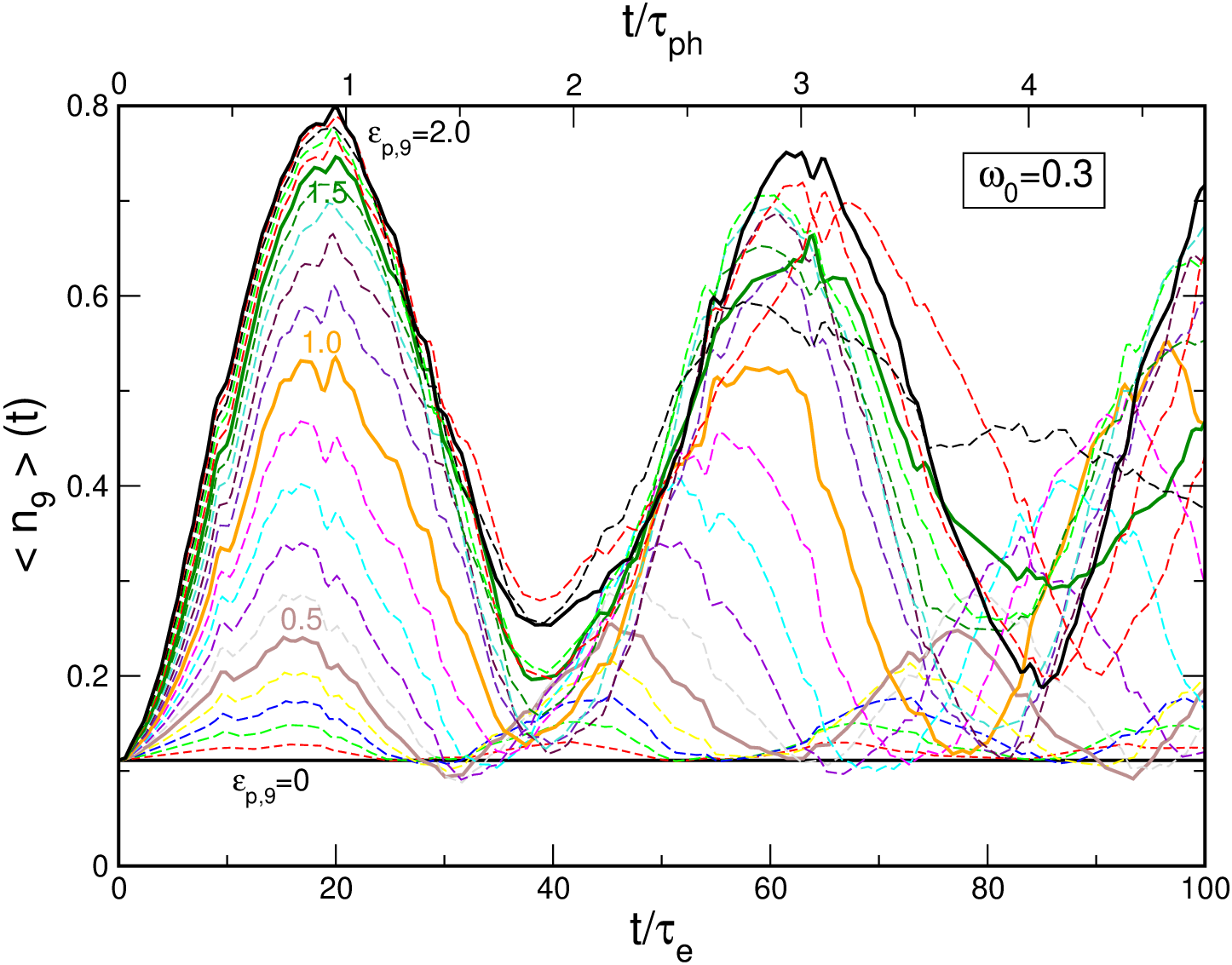}
\includegraphics[width=.9\linewidth]{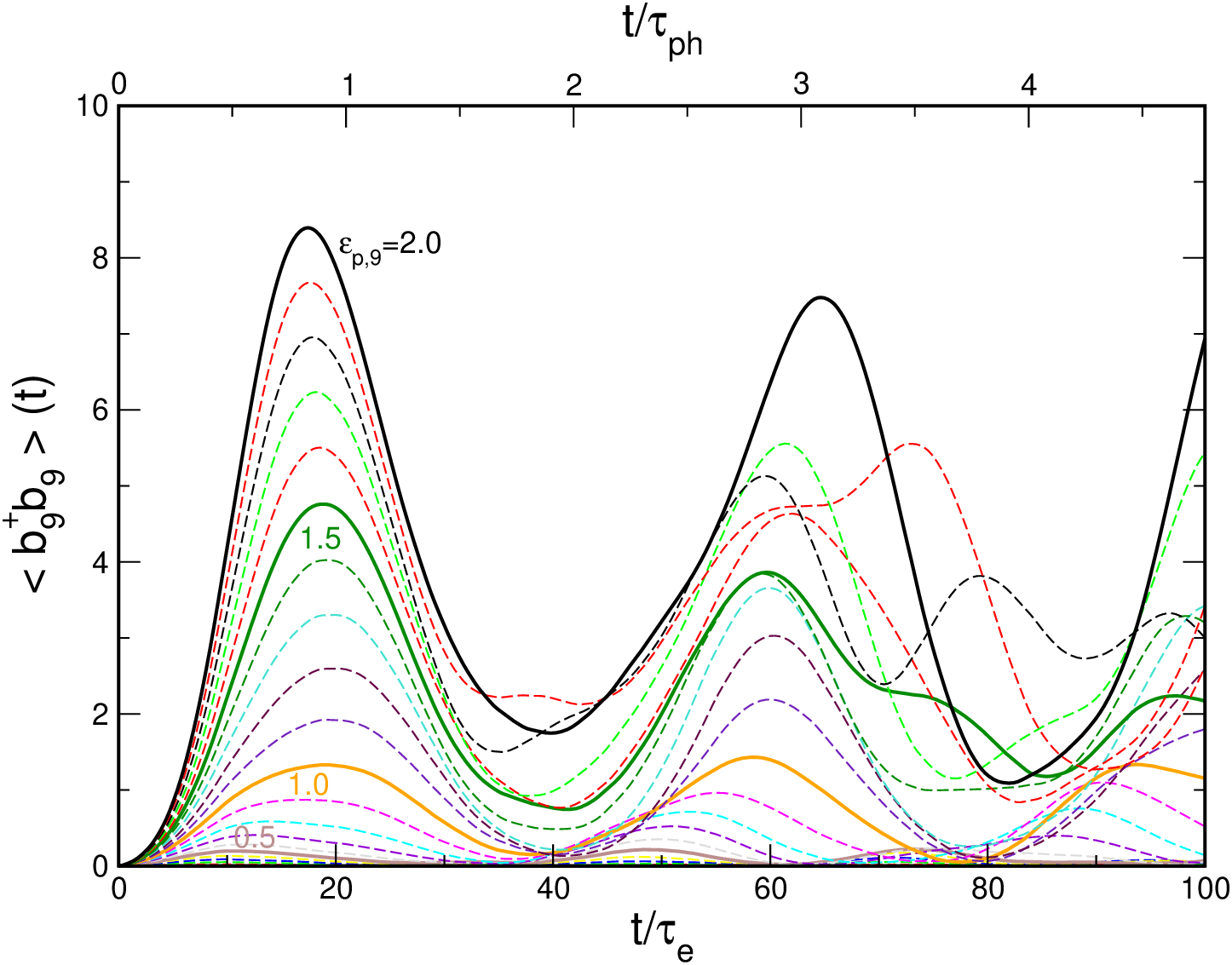}
\caption{(Color online)  Time dependence of the particle density (upper panel) and phonon number (lower panel)
at the deformable ``molecular crystal'' site 9,  starting out from the free-electron ground state of a 17 site chain with OBCs.  The phonon frequency $\omega_0=0.3$.
The EP coupling $\varepsilon_{p,9}$ is switched on at $t=0$, different curves belong to $\varepsilon_{p,9}$ values  increased by 0.1. In the numerical calculations we take into account up to  $M=100$ phonons,  $N_{\rm C}=10$ Chebyshev moments, and use a time step $\Delta t/\tau_e=0.01$.}
\label{fig:n_b_w0.3_9_wc}
\end{center}
\end{figure}

\begin{figure}[htbp]
\begin{center}
\includegraphics[width=0.95\linewidth]{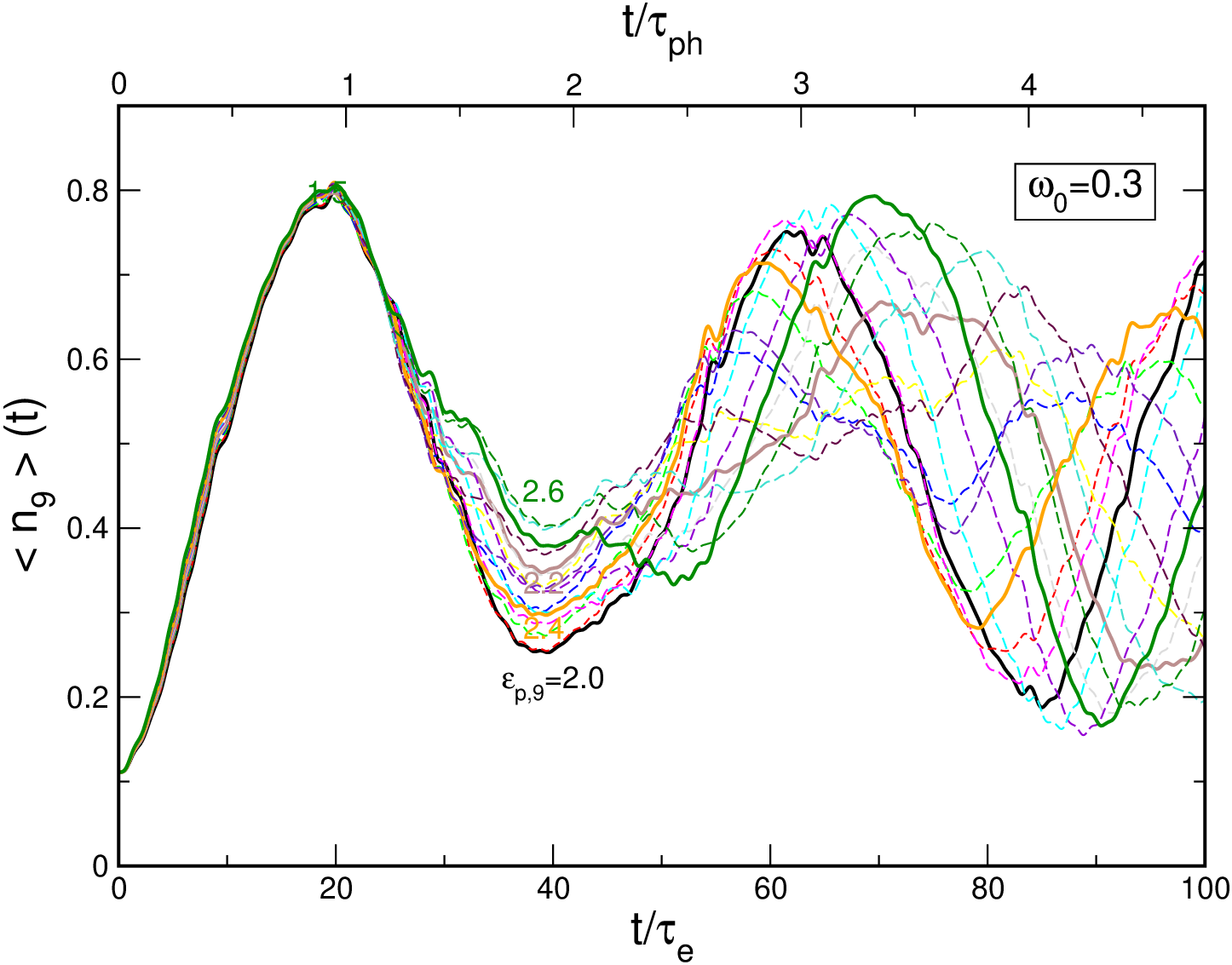}
\includegraphics[width=0.95\linewidth]{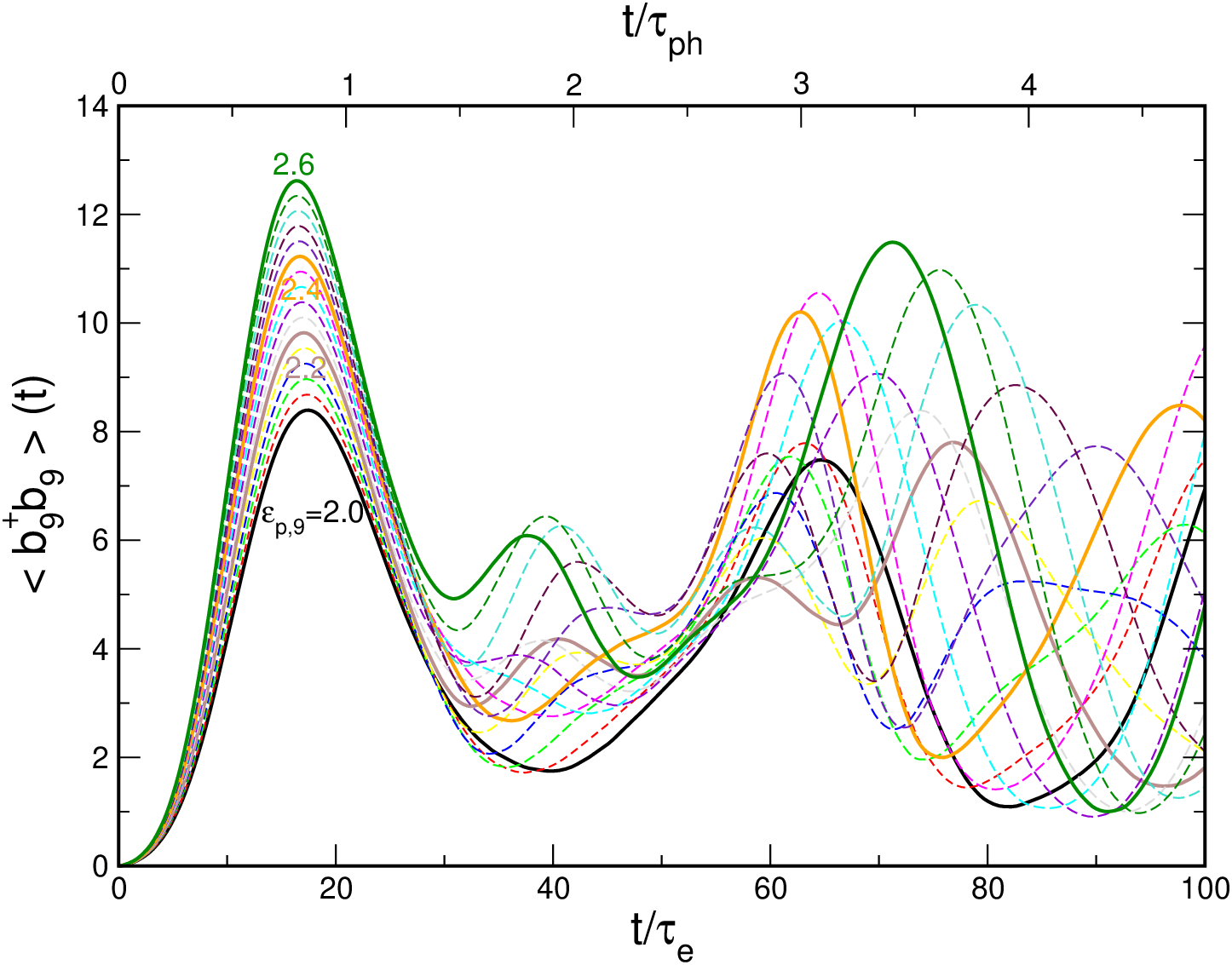}
\caption{(Color online) Time dependence of the particle density (upper panel) and phonon number (lower panel)
at molecular crystal site 9 in the intermediate EP coupling adiabatic regime. The initial state is the same 
as in Fig.~\ref{fig:n_b_w0.3_9_wc}; $\varepsilon_{p,9}$ now is increased in steps of 0.04.}
\label{fig:n_b_w0.3_9_ic}
\end{center}
\end{figure}

Figure~\ref{fig:n_b_w0.3_9_wc} illustrates the time evolution of the particle density at the oscillatory site
after the interaction quench for weak-to-intermediate EP couplings and phonon frequencies $\omega_0$ smaller 
than the electronic transfer integral $t_0$. We note that the electron
is not uniformly spread over the lattice even at $t<0$ where  $\varepsilon_{p,9}=0$ because of of the OBCs:  
$\langle n_9\rangle (0) =1/9$ is roughly twice the mean electron density 1/17. 

After  the local EP 
interaction is turned on the electron can couple to the molecular vibrations at site~9. The basic interaction 
process is the absorption and emission of a phonon by the electron with a simultaneously change of the 
electron state.  At the same time the lattice is distorted locally.   Such a lattice distortion may trap
the charge carrier if the EP coupling is strong. As a result the local particle density is enhanced. 
Since the trapping potential itself 
depends on the carrier's state, this highly non-linear feedback phenomenon is 
called ``self-trapping''.~\cite{Fi75,WF98a} 
Figure~\ref{fig:n_b_w0.3_9_wc} clearly shows an initial strong increase of the local particle
density in time. Since the characteristic nearest-neighbor hopping time of a bare electron is $\tau_e$,  
all ``electrons'' initially  moving toward the central site will reach this site within $t/\tau_e\leq 8$ 
(dealing with a single particle we actually thereby think of electronic contributions). 
This explains the small hump on the left shoulder of the first $\langle n_9\rangle$--increase at about $t/\tau_e\sim 8$. Electrons that move away from the central 
site will reach it after reflection at the boundaries within  the time interval $8<t/\tau_e\leq 17$.  
So if the particle is held at the molecular site by the EP coupling, it is trapped to the greatest 
possible extent at about  $t/\tau_e\sim 17$ (which almost coincides with  the first 
maximum in $\langle n_9\rangle$ at $t_{\rm max}^{(1)}/\tau_e\simeq 20$). 
Self-evidently the maximum is enhanced as $\varepsilon_{p,9}$
increases. 

When an electron reaches site~9 it can emit a phonon to lower its energy. 
The phonon period is $\tau_{ph}$. Hence, for the adiabatic regime discussed in 
Fig.~\ref{fig:n_b_w0.3_9_wc}, the phonon excited by the first arriving ``part of the electron'' 
is still present when the last part of the electron arrives (cf. the phonon time scale
displayed in the graphs at the opposite $x$-axes). Because of this retardation effect 
the number of phonons at the molecular site~9  steadily increases and 
$\langle b_9^\dagger b_9^{}\rangle~$ develops, for $\omega_0=0.3$, 
a maximum in time slightly before $\langle n_9\rangle$ reaches its maximum.
As time proceeds further, the particle starts hopping further  away from
the oscillatory site (recall that the system is no longer in an eigenstate after the
interaction quench) and $\langle n_9\rangle (t)$ decreases until the whole process
recurs. Importantly, the particle density $\langle n_9\rangle (t)$ at  its first mimium 
around $t_{\rm min}^{(1)} \sim  2 t_{\rm max}^{(1)}\simeq 40 \tau_e$  
is substantially larger than at $t=0$ above
the ``critical'' EP coupling $\varepsilon_{p,9}/2t_0\simeq 1$. This gives a first indication that indeed a 
polaron is formed at site~9, in contrast to the weak EP coupling case.
\begin{figure}[t]
\begin{center}
\includegraphics[width=0.95\linewidth]{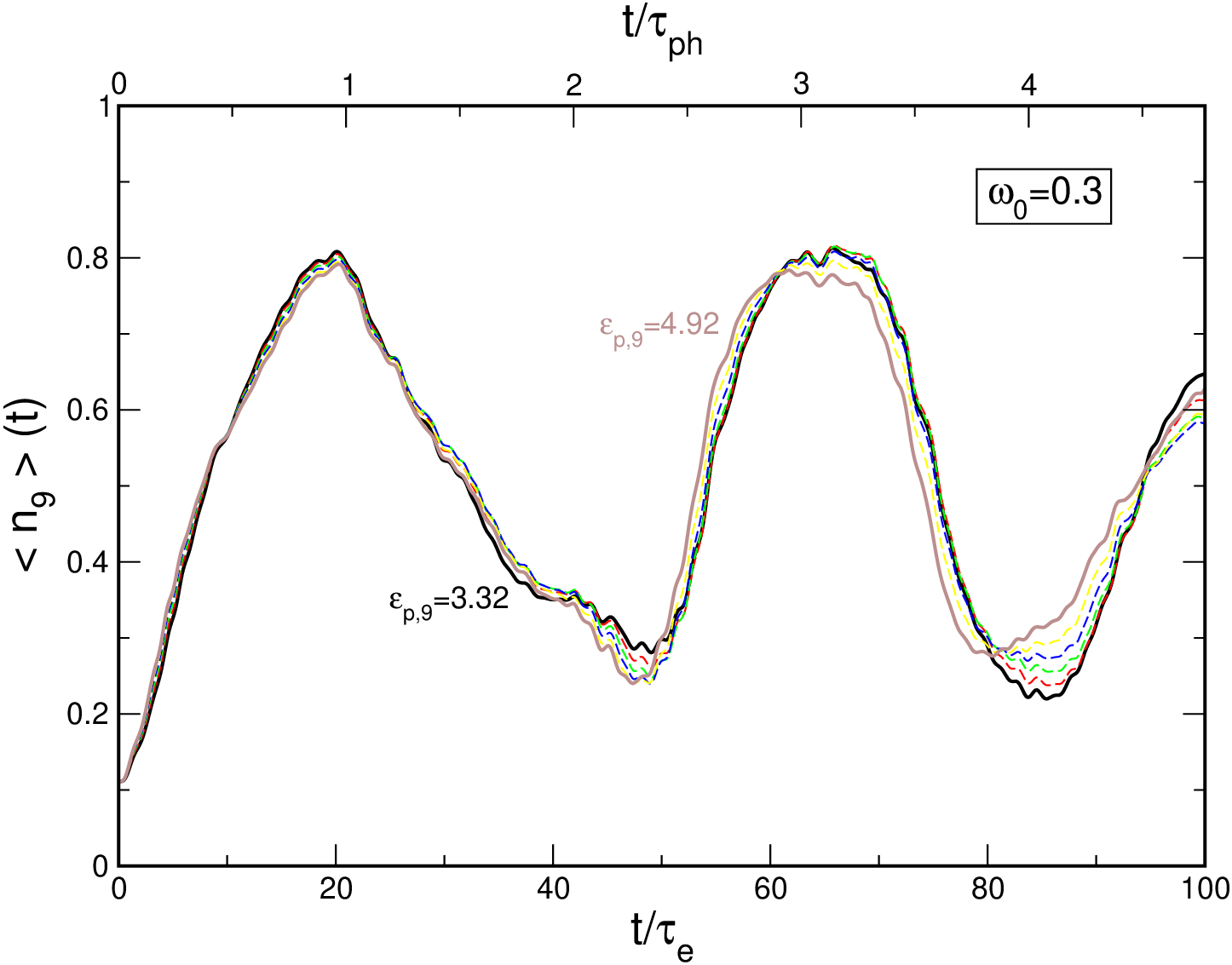}
\includegraphics[width=0.95\linewidth]{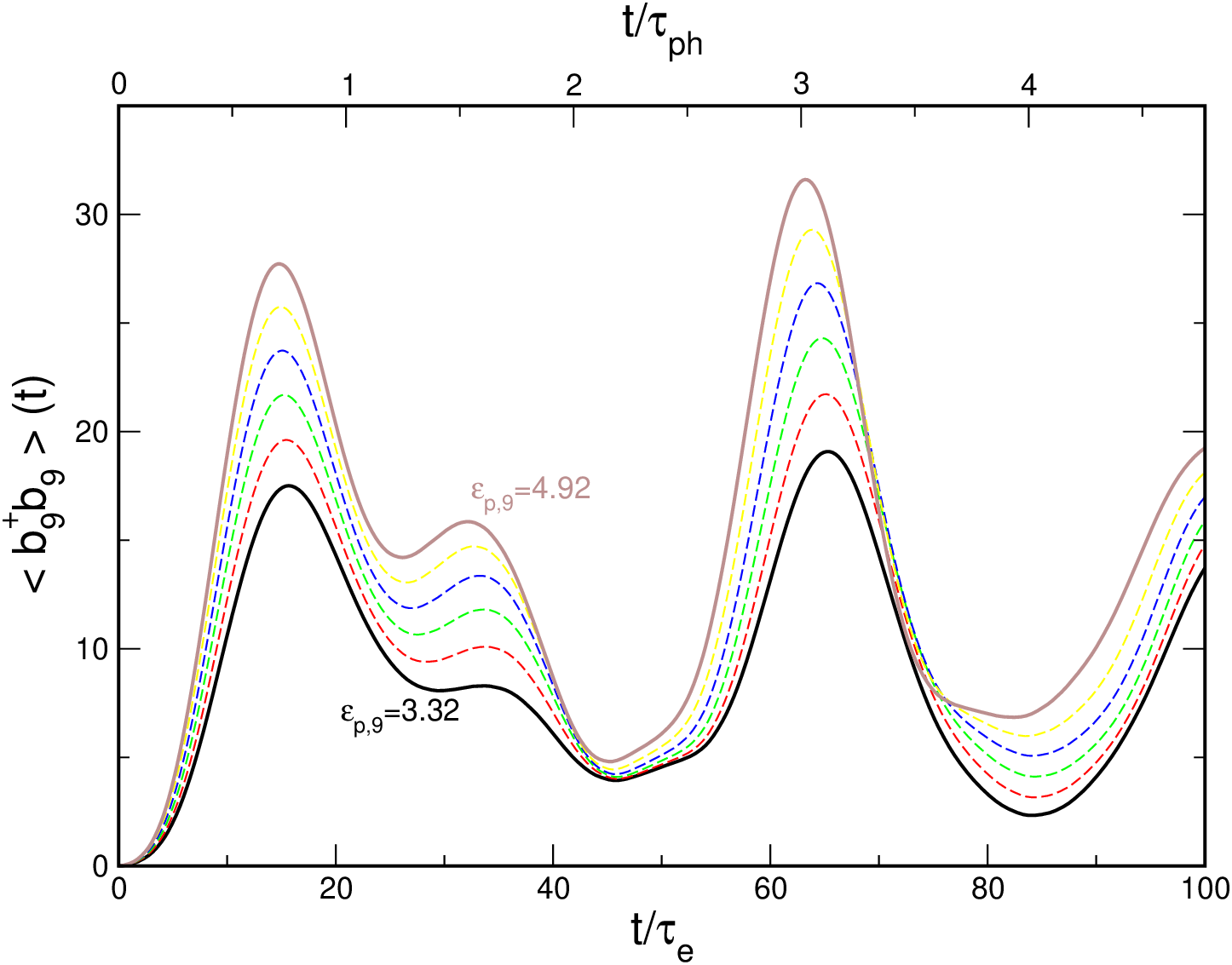}
\caption{(Color online) Time dependence of the particle density (upper panel) and phonon number (lower panel)
at site 9 in the strong  EP coupling adiabatic regime. Initial state 
as in Fig.~\ref{fig:n_b_w0.3_9_wc}; $\varepsilon_{p,9}$ is increased in steps of 0.32.}
\label{fig:n_b_w0.3_9_sc}
\end{center}
\end{figure}

Figures~\ref{fig:n_b_w0.3_9_ic} and~\ref{fig:n_b_w0.3_9_sc}
demonstrate that at larger EP couplings the particle density at
the oscillatory site~9 evolves in almost the same manner
for a relatively long time span. This especially holds for the 
strong-coupling case displayed in Fig.~\ref{fig:n_b_w0.3_9_sc}, 
where every incoming electron sticks to the molecular site. 
Thereby the total phonon number increases with increasing $\varepsilon_{p,9}$. Then the  interesting question
is, of course, whether (or to what extent) the excited phonons are incorporated in the polaronic 
quasiparticle or rather will be uncorrelated. In our case, where the EP coupling acts on a single site
only, the electron can not carry a phonon cloud away (this--more realistic--situation will be investigated
in the subsequent two sections). 
Nevertheless we can address this question by analyzing the phonon 
distribution function, $|c_m|^2(t)$ with $\sum_{m=0}^M |c_m|^2(t)=1$, yielding the weight 
of the $m$-phonon contribution in the wave function $|\psi(t)\rangle$.~\cite{JF07}

\begin{figure}[t]
\begin{center}
\includegraphics[width=0.95\linewidth]{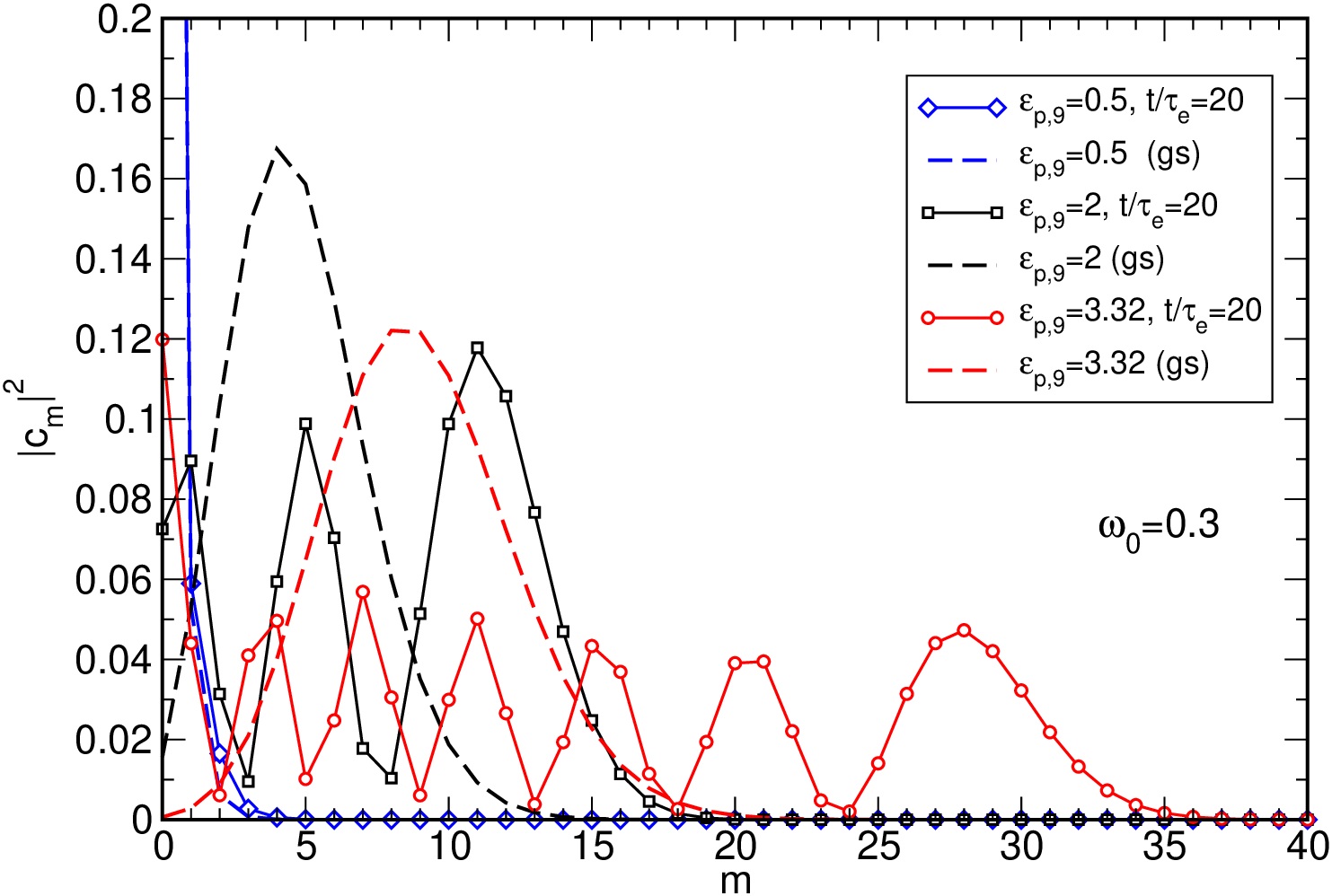}
\includegraphics[width=0.95\linewidth]{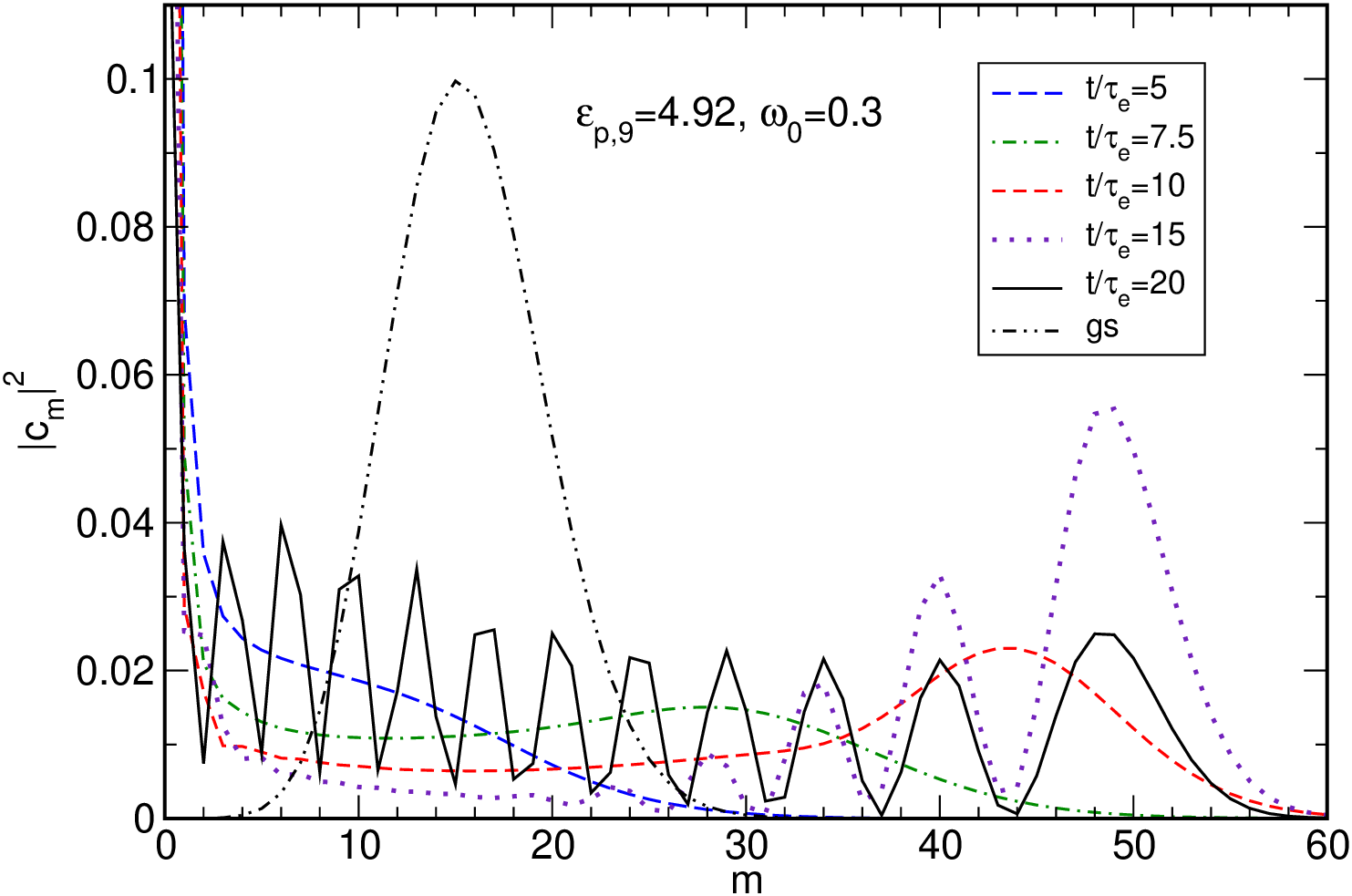}
\caption{(Color online)  Contribution of the $m$-phonon state  to $|\psi(t)\rangle$ at different
times $t$ (given in units $\tau_e$). Results for  various $\varepsilon_{p,9}$ were 
compared to the phonon distribution function of the system's ground state
 ($\varepsilon_{p,9}>0\forall\,t$, no interaction quench). The phonon frequency is $\omega_0=0.3$.}
\label{fig:pdf_w0.3}
\end{center}
\end{figure}

Figure~\ref{fig:pdf_w0.3} gives $|c_m|^2(t)$ for different EP interaction strengths, ranging
from weak to strong couplings. For comparison, the corresponding phonon distribution 
functions of the stationary ground states (where $\varepsilon_{p,9}>0$ $\forall t$) is shown.
At small $\varepsilon_{p,9}$, $|\psi(t)\rangle$ basically is a zero-phonon state at any time.
As a matter of course a few phonons will be emitted but immediately after will be reabsorbed as the particle passes
the  molecular site. Therefore, no long-living lattice distortion appears that might trap the carrier.
The situation dramatically changes as $\varepsilon_{p,9}$ exceeds the critical coupling strength
for polaron formation. Now the phonon distribution of the ground state is Poisson distributed with maxima
at about 4 ($\varepsilon_{p,9}=2$), 9 ($\varepsilon_{p,9}=3.32$), and 15 ($\varepsilon_{p,9}=4.92$). 
$|\psi(t=20\tau_e)\rangle$ is a multi-phonon state as well. Since $\omega_0$ is rather small,  
an adiabatic potential (energy) surface emerges 
that retains the incoming electron contributions so that the formation of an adiabatic Holstein polaron~\cite{Ho59a,WF98a}
can occur. The initial energy of our system ($E_0=-1.9696$), however, does not allow the particle to access the polaronic
ground state having $E_0(\varepsilon_{p,9}=2)=-2.521$, $E_0(3.32)=-3.628$, and $E_0(4.92)=-5.126$.
As can be seen from Fig.~\ref{fig:pdf_w0.3},  the form of the phonon distribution function reflects the phonon distribution of excited 
displaced harmonic oscillator states, indicating that excited states of the polaron were realized instead. 
These states are known to be separated in energy  by $\omega_0$.~\cite{FT07}  
Indeed, in going e.g. from $\varepsilon_{p,9}=3.32$ to $\varepsilon_{p,9}=4.92$ five additional 
phonons were created (all bound to the polaron), giving rise to a polaron excited state.   
Note that such kinds of phonon distributions were found for  the Raman- and infrared-active intrinsic
localized modes in quasi one-dimensional mixed-valence transition-metal 
complexes.~\cite{Swea99}
\begin{figure}[htbp]
\begin{center}
\includegraphics[width=0.95\linewidth]{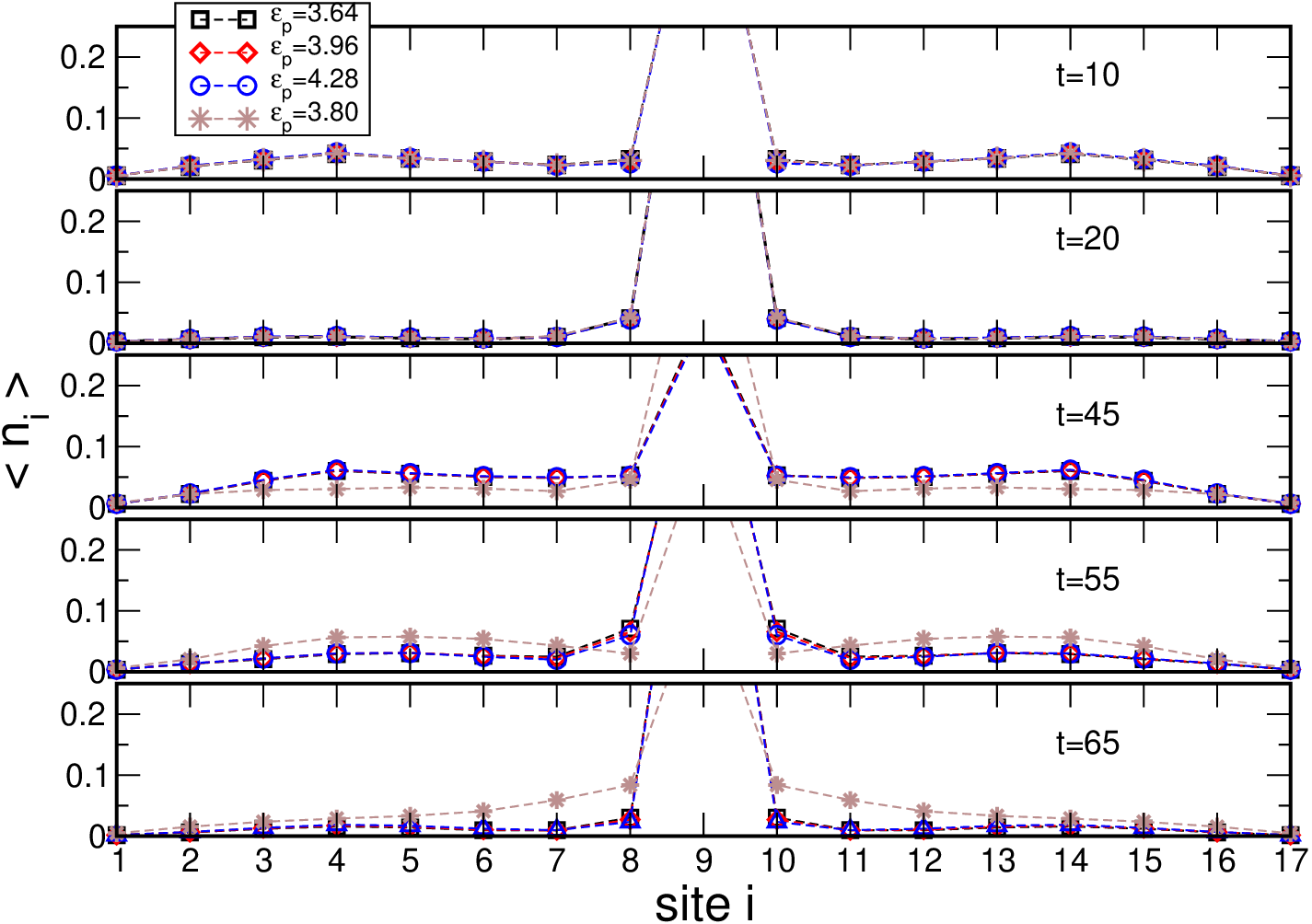}
\caption{(Color online) Electron densities on the one-dimensional chain at different times $t/\tau_e$.
Initial state as before; $\omega_0=0.3$.}
\label{fig:ni_w0.3_sc}
\end{center}
\end{figure}

The lower panel of Fig.~\ref{fig:pdf_w0.3} yields some insight on the time scale  that 
the polaron formation process takes place. Up to $t/\tau_e=10$ the electron radiates
successive phonons which are still uncorrelated, however. Therefore all phonon states below a certain 
threshold are equally well represented in $|\psi(t)\rangle$.  The zero-phonon
state has a larger weight, of course, because parts of the electron still reside outside the molecular site.  
At about $t/\tau_e=15$ the phonons become correlated, i.e. they are increasingly tightly 
bound to the electron. This process is completed at  $t/\tau_e\sim 20$.

In Fig.~\ref{fig:ni_w0.3_sc} we show snapshots of the electron density 
distribution along the whole chain at various points in time.  
We see that by increasing the EP interaction in such a way that 
the energy of the initial state matches one of the polaron excited 
states (starting out from the well-established polaronic state at 
$\varepsilon_{p,9}=3.32$, see Figs~\ref{fig:n_b_w0.3_9_sc} 
and~\ref{fig:pdf_w0.3}), the spatiotemporal variation of the 
various excited states is the same, even for very long times and away from 
the central molecular site (cf. the curves marked by squares, diamonds and circles for $\varepsilon_{p,9}=3.64$, 3.96 and~4.28, respectively). 
In contrast, if we choose  an EP coupling which does not match the 
ground-state energy $E_0$ by lowering the polaron-level ladder, 
after a while the densities evolve quite differently [cf. data for
$\varepsilon_{p,9}=3.80$ (stars)].  

\subsubsection{Anti-adiabatic regime} 
We next investigate the limit of large phonon frequencies. Now $\tau_{ph}$ is comparable or even
smaller than $\tau_e$, which means that a phonon can be excited and re-absorbed instantaneously 
when the electron enters the molecular site. During this process the electron will become (partly) dressed 
by phonons, provided the EP interaction is sufficiently strong. In this case a non-adiabatic 
Lang-Firsov-type polaron is formed.~\cite{LF62,WF98a} This will not happen in the weak EP coupling regime 
illustrated in Fig.~\ref{fig:n_b_w8_9_wc}.   
Because of the extremely large phonon energy, only very few phonons can be radiated by the electron
(see lower panel). The molecular-site particle density shown in the upper panel is weakly modulated
by the phonon emission-absorption processes on the time scale $\tau_{ph}$ and ``oscillates'' on a 
time scale related to the system size $N=17$ (within $t/\tau_e=17$ all electrons have visited the central
site once). No retardation phenomena are observed (in  contrast to Fig.~\ref{fig:n_b_w0.3_9_wc}).
The dip around  $t/\tau_e=10$ is because electrons continously leaving the dot and those arriving at this time 
mostly come from the boundary where the electron density at $t=0$ was very small. The situation 
becomes more complex as $\varepsilon_{p,9}$ increases and the electron at the molecular
site will be partly dressed leading to stronger fluctuations of the phonon number.
\begin{figure}[htbp]
\begin{center}
\includegraphics[width=0.95\linewidth]{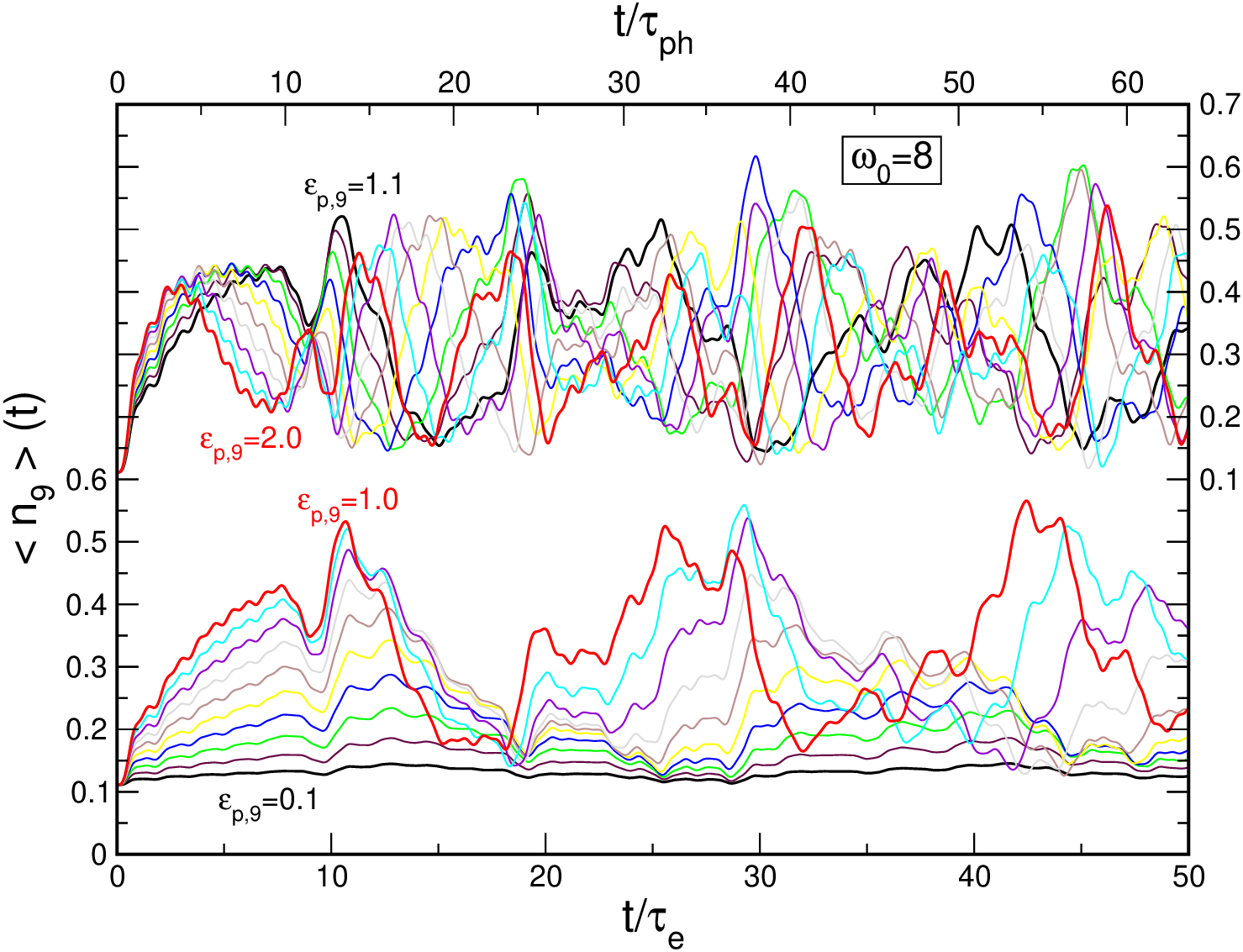}
\includegraphics[width=0.95\linewidth]{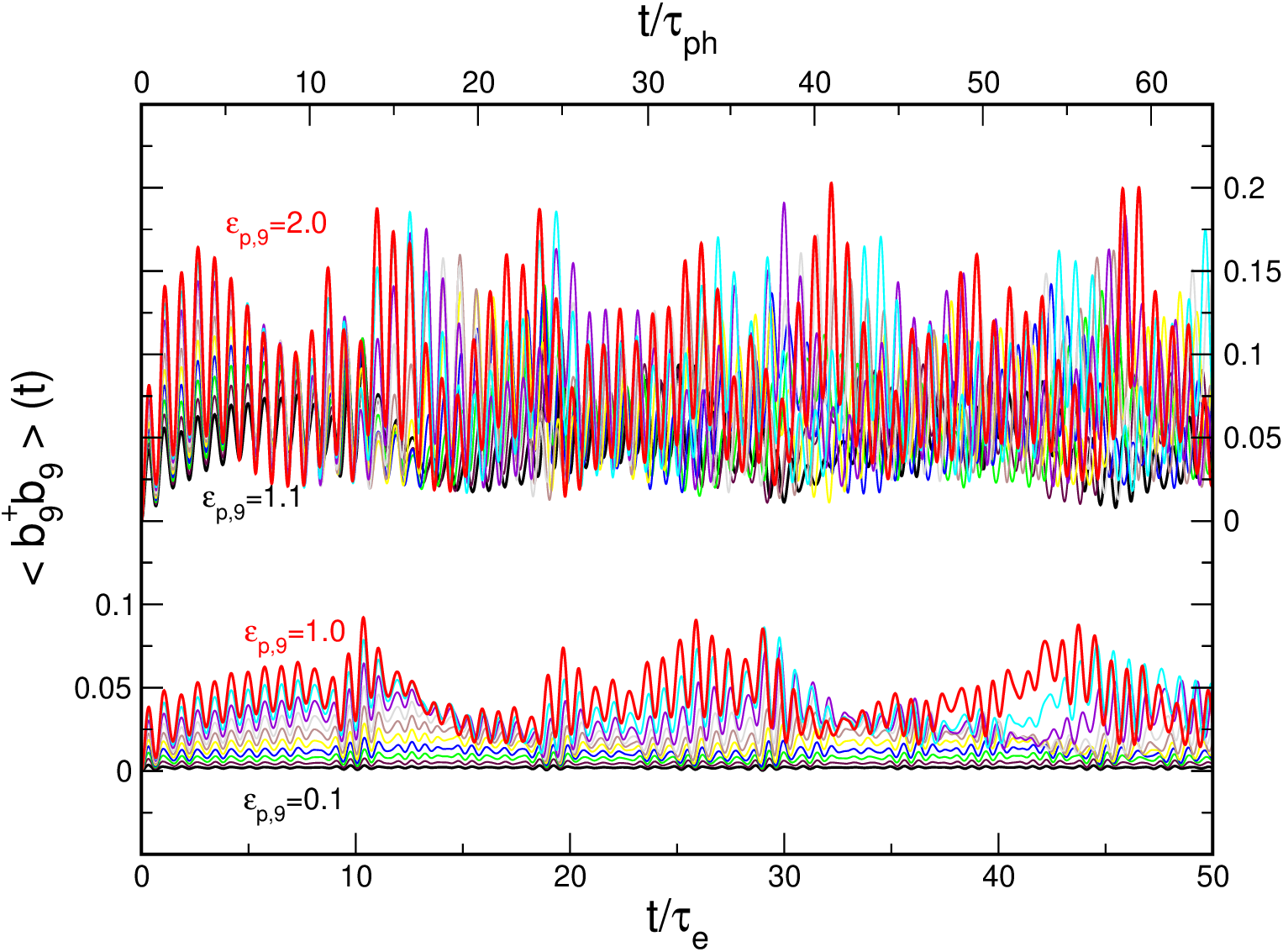}
\caption{(Color online) Time dependence of the particle density (upper panel) and phonon number (lower panel)
at molecular crystal site 9 in the weak EP coupling anti-adiabatic  regime. The phonon frequency  $\omega_0=8$.
The EP coupling is switched on at $t=0$; different curves belong to $\varepsilon_{p,9}$ values 
increased by 0.1.}
\label{fig:n_b_w8_9_wc}
\end{center}
\end{figure}

At ``intermediate'' couplings  $\varepsilon_{p,9}=10$ (note that $\varepsilon_{p,9}=10/\omega_0$ is of the order of one),
the ground-state energy $E_0(10)=-10.1215$ and the system comes into ``resonance'' with the initial state (having  $E_0\sim -2$) by exciting just one phonon (see Fig.~\ref{fig:n_b_w8_9_ic}). Since $g_9^2 > 1$ a polaron 
will evolve, i.e. the phonon is bound to the electron. The same happens at  $\varepsilon_{p,9}=18$
($E_0(18)=-18.0622$), but now two phonons will be excited and incorporated. There are two points worth mentioning.
First, as the system oscillates on its $t/\tau_e=17$-period it can adjust far better in order to  form a polaron
at the sequent $t_{\rm max}$-points. Second, putting $\varepsilon_{p,9}$ only somewhat out of tune,
both $\langle n_9\rangle$ and  $\langle b_9^\dagger b^{}_9\rangle$ are substantially reduced for all $t$,
i.e. polaron formation is suppressed. 

\begin{figure}[htbp]
\begin{center}
\includegraphics[width=0.95\linewidth]{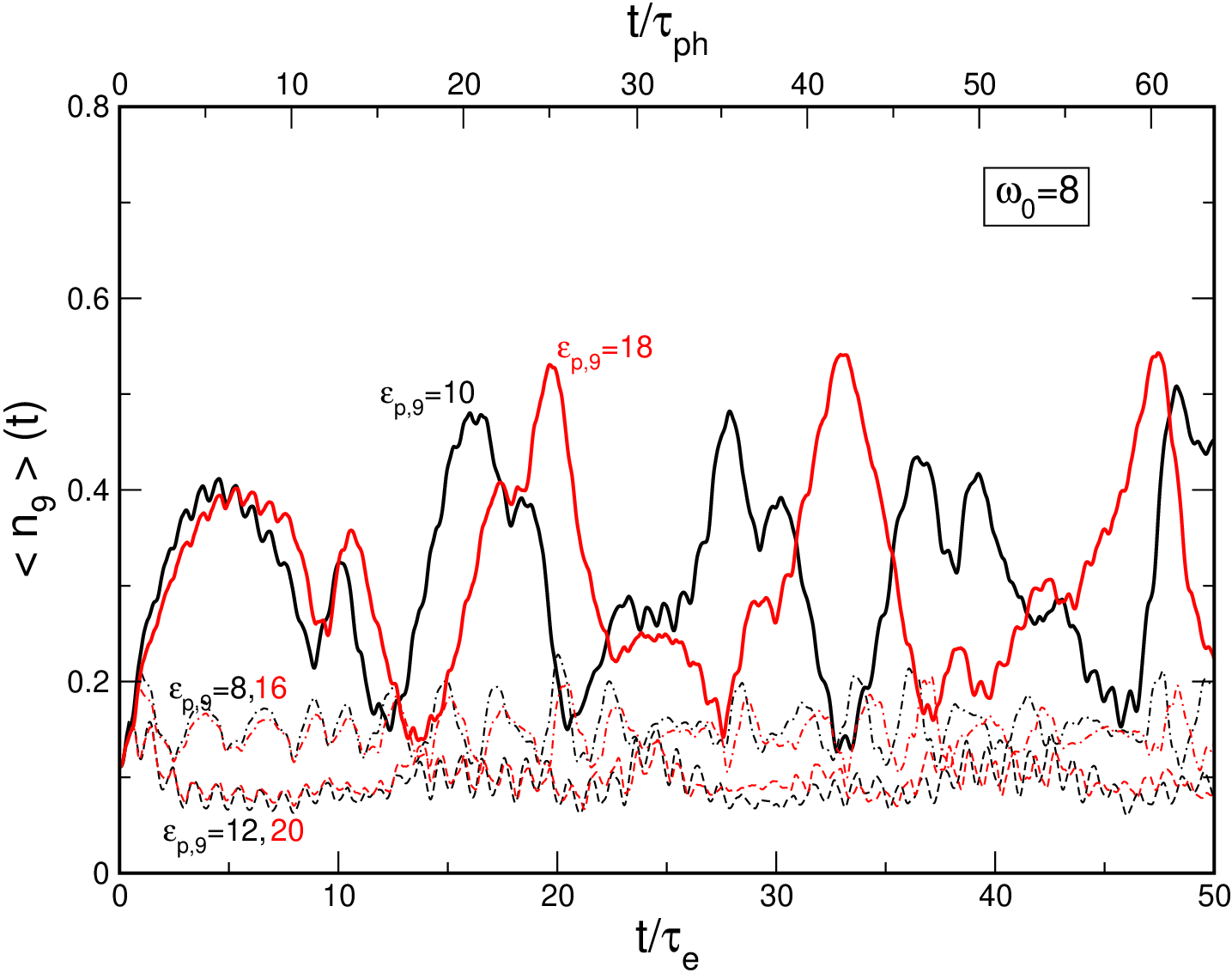}
\includegraphics[width=0.95\linewidth]{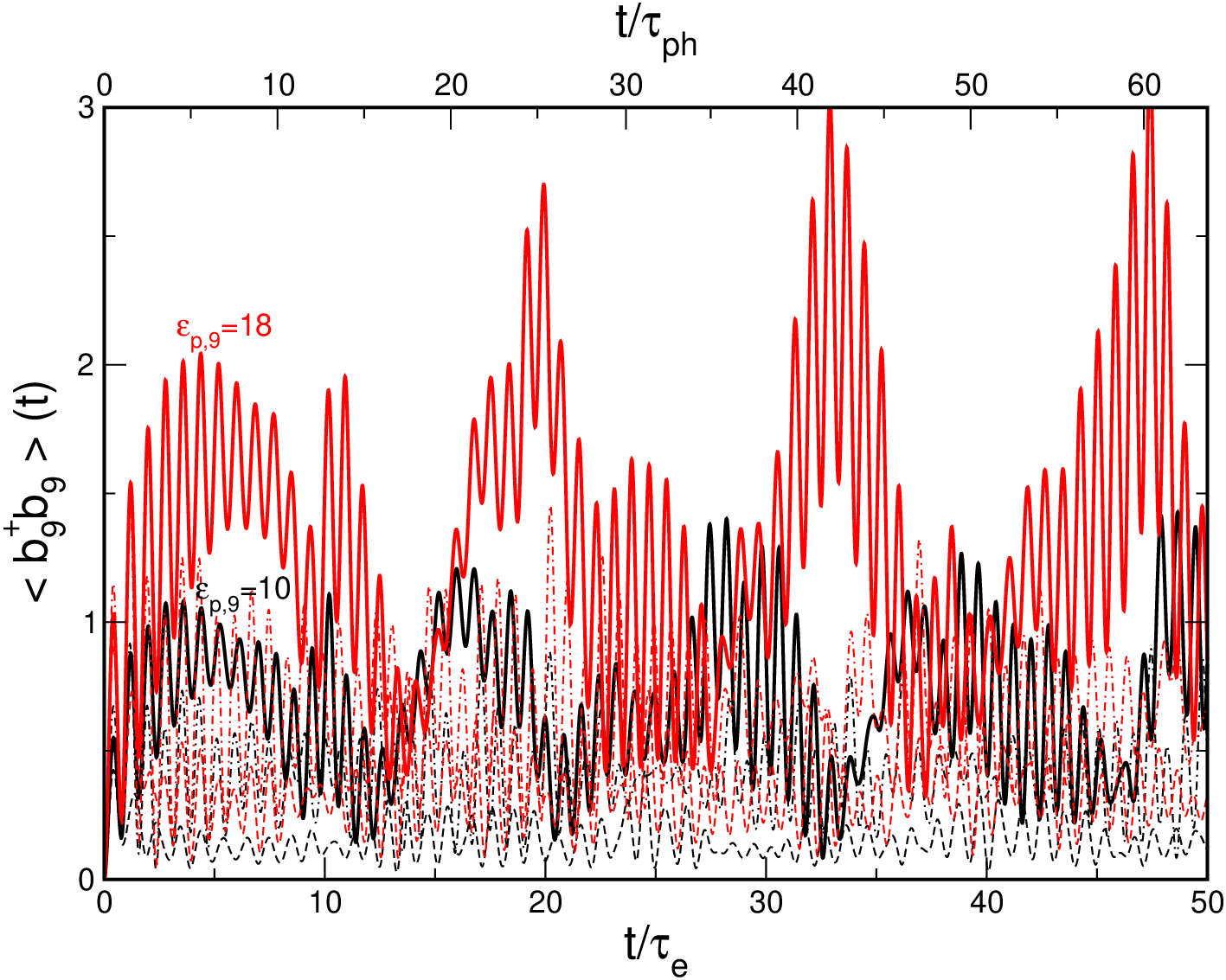}
\caption{(Color online) Time dependence of the particle density (upper panel) and phonon number (lower panel)
at the molecular crystal site 9 in the strong  EP coupling anti-adiabatic regime. }
\label{fig:n_b_w8_9_ic}
\end{center}
\end{figure}

In the strong EP coupling regime displayed in Fig.~\ref{fig:n_b_w8_9_sc}, the anti-adiabatic Lang-Firsov
polaron has been fully developed. In this case the arriving electronic contributions stay at the 
molecular site for such a long time that $\langle n_9 \rangle$ reaches 0.8. Note the increase of 
$t_{\rm max}^{(1)}$ as compared to Figs.~\ref{fig:n_b_w8_9_wc} and~\ref{fig:n_b_w8_9_ic} 
(in order to demonstrate that $t_{\rm max}^{(1)}$ is indeed determined by the system size we 
included results for a system with $N=21$ sites).  The lower panel makes clear that now 
on average many more phonons ($\propto g_9^2$) were incorporated.

\begin{figure}[htbp]
\begin{center}
\includegraphics[width=0.95\linewidth]{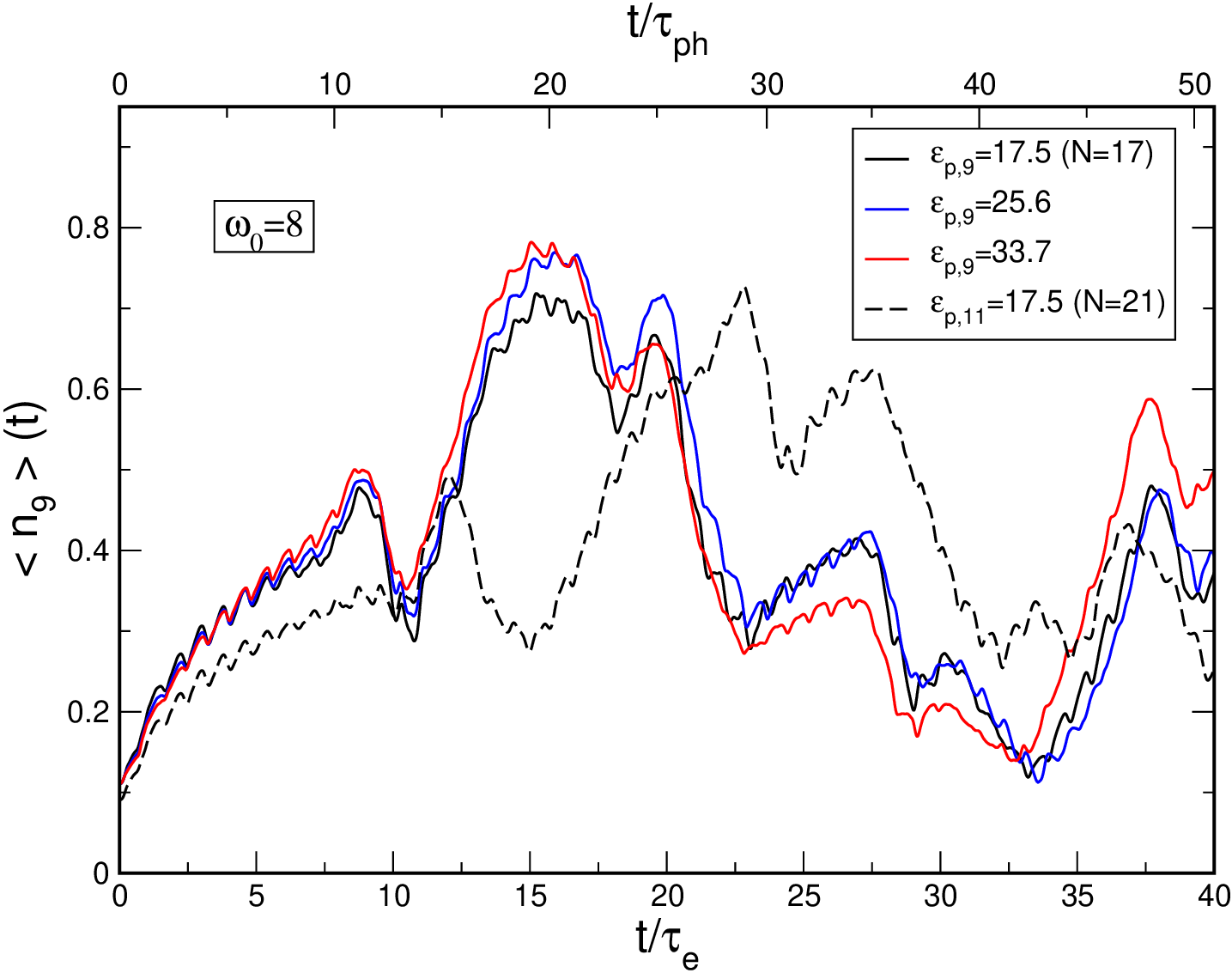}
\includegraphics[width=0.95\linewidth]{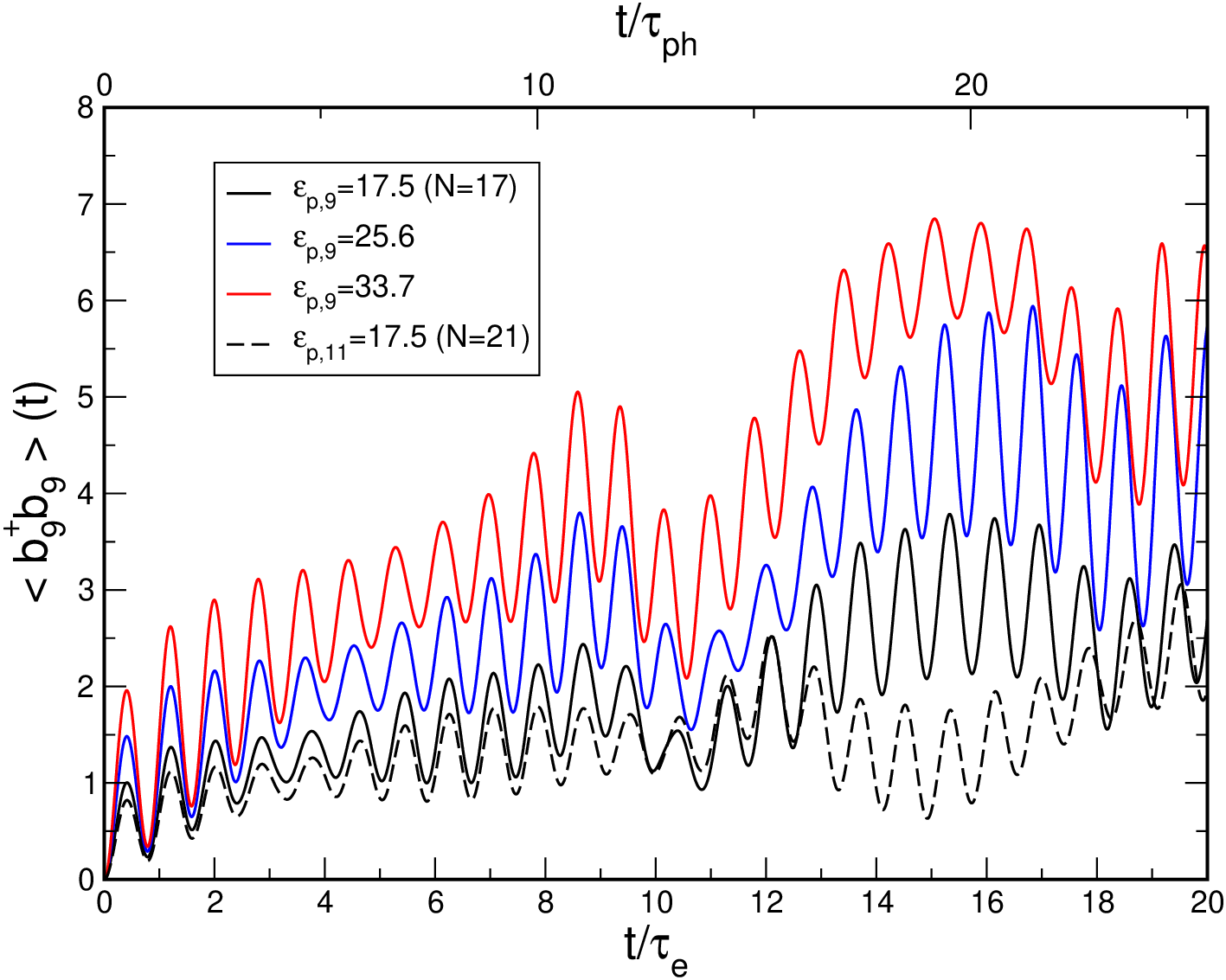}
\caption{(Color online) Time dependence of the particle density (upper panel) and phonon number (lower panel)
at the molecular crystal site 9 in the extremely strong  EP coupling anti-adiabatic regime. The EP interaction is increased 
in steps of $\omega_0$. For $N=21$ the particle and phonon numbers are displayed at site~11.}
\label{fig:n_b_w8_9_sc}
\end{center}
\end{figure}

The phonon distribution function shown in Fig.~\ref{fig:pdf_w80} corroborates this scenario.
As for the adiabatic case (cf.  Fig.~\ref{fig:pdf_w0.3}), we observe a transition from
an uncorrelated few-phonon state to a correlated multi-phonon polaron state.
The phonon distribution function shows that $|\psi(t^{(1)}_{\rm max})\rangle$  corresponds to an excited polaron
with  two (four) bound phonons at $\varepsilon_{p,9}=18$ ($\varepsilon_{p,9}=33.7$).

\begin{figure}[htbp]
\begin{center}
\includegraphics[width=0.95\linewidth]{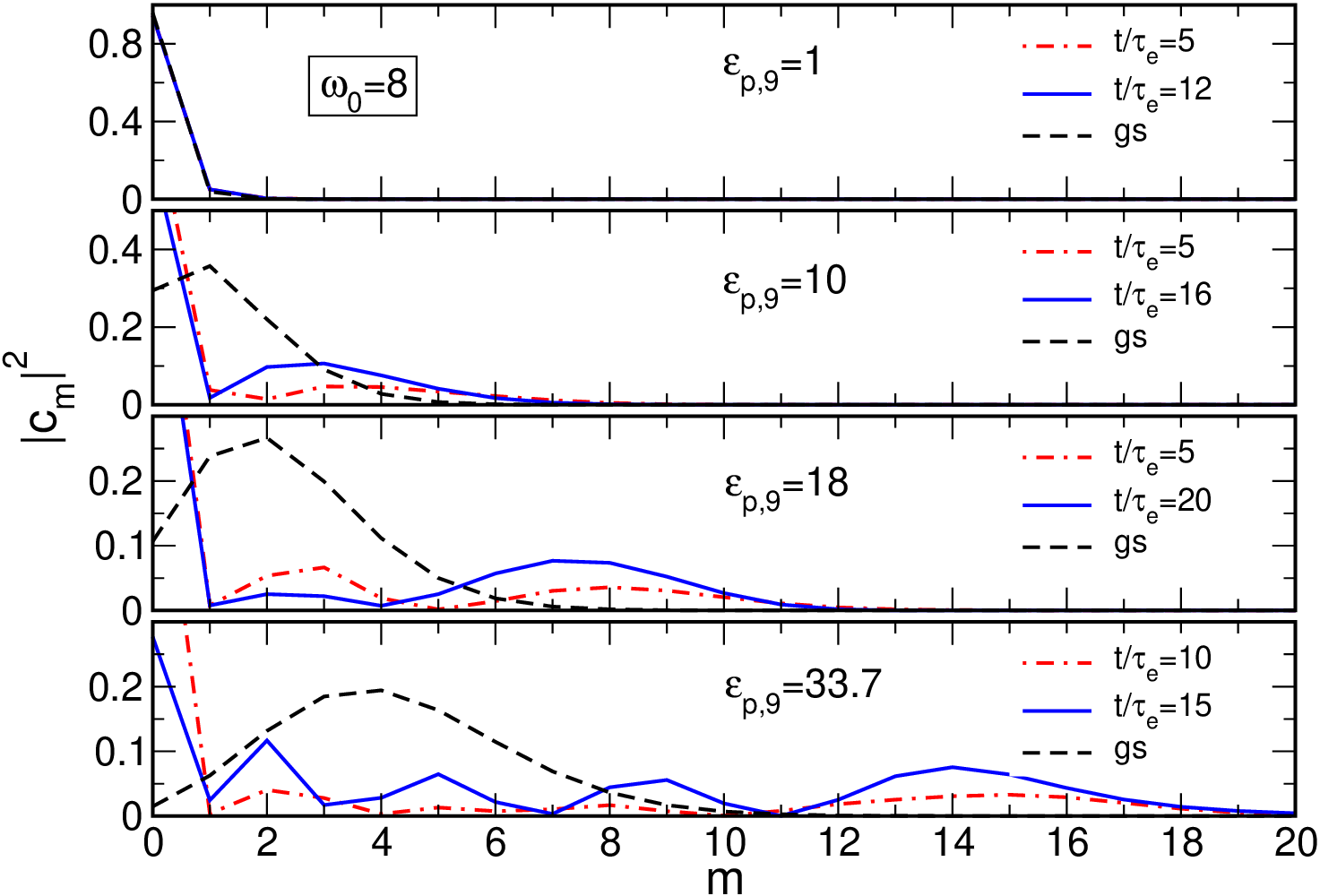}
\caption{(Color online) Contribution of $m$-phonon states  to $|\psi(t)\rangle$ at different
EP couplings and points of time. Results were 
compared to the stationary phonon distribution function of the system's true ground state. 
The phonon frequency is $\omega_0=8$.}
\label{fig:pdf_w80}
\end{center}
\end{figure}

\subsection{Wave-packet injection}
A bare electron injected into a quantum wire coupled to the lattice vibrations
at every lattice site is another instructive example for the quantum dynamics
of polaron formation.~\cite{KT07} To this end, at $t=0$, we place a Gaussian
wave packet of width $\sigma_0^2=3/\ln 10$ and momentum $K$ centered at site
$l_0=4$,
\begin{equation}
|\psi(0)\rangle= A \sum_{l=1}^7 {\rm e}^{\big[-\tfrac{(l-l_0)^2}{2\sigma_0^2}\big]}
{\rm e}^{{\rm i}K(l-l_0)}
c_l^\dagger |0\rangle\,,
\label{wave_packet}
\end{equation}
and let it evolve ($A$ is a normalization constant to ensure $\langle\psi(0)|\psi(0)\rangle =1$). Again $\Delta_i=0$ $\forall i$.
\begin{figure}[htbp]
\begin{center}
\includegraphics[width=0.95\linewidth]{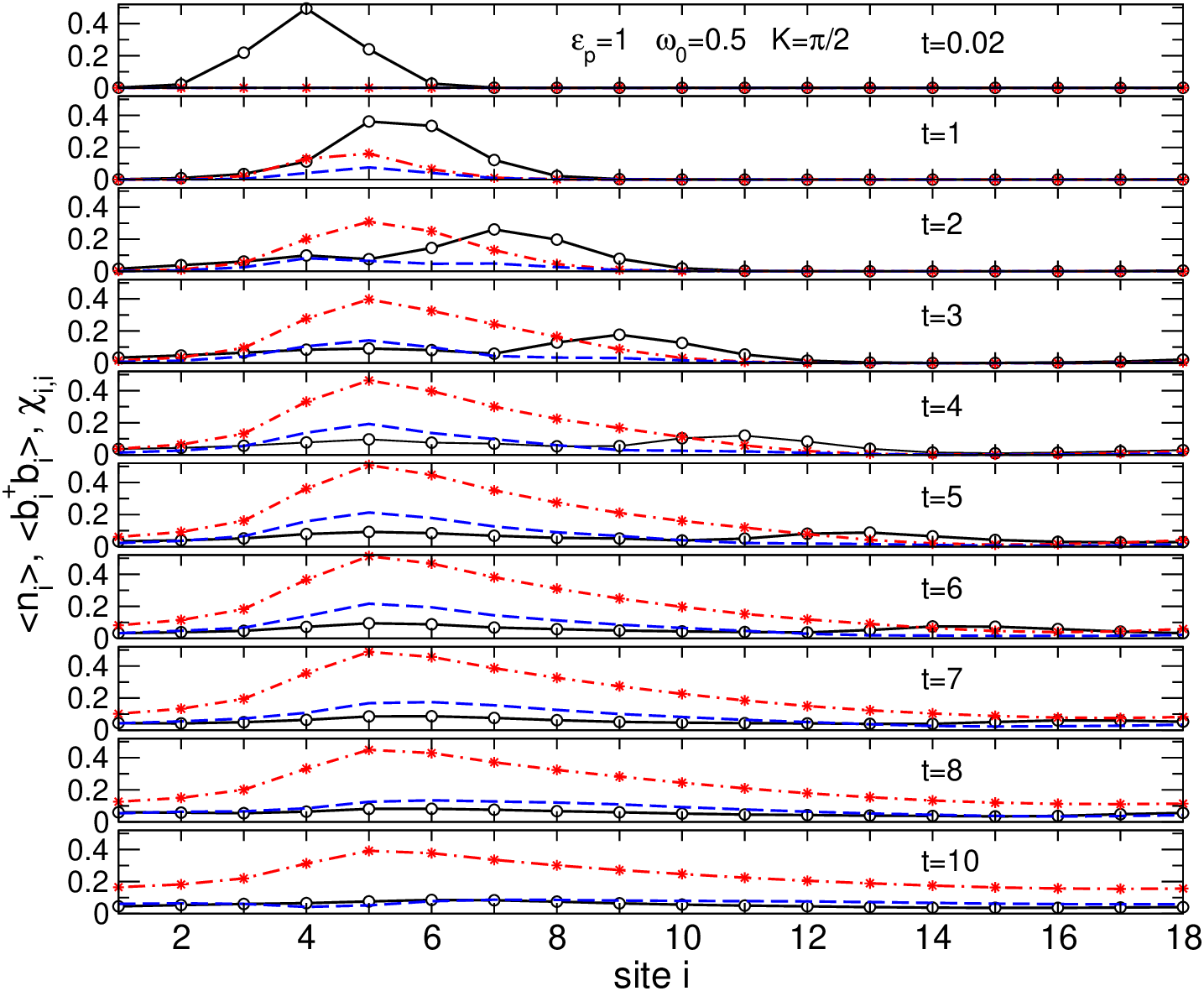}
\caption{(Color online) Spatiotemporal evolution of a free electron Gaussian wave packet 
injected with wave vector $K=\pi/2$ at $t=0$ 
into an 18-site molecular crystal chain with PBCs. 
Model parameters are $\varepsilon_p=1$ and $\omega_0=0.5$.
Displayed is the time evolution (in units of $\tau_e$) of the local particle densities $\langle n_i\rangle$ 
(black solid lines, open circles), phonon numbers $\langle b_i^\dagger b_i^{}\rangle$
(red dot-dashed lines, stars) and local EP correlations $\chi_{i,i}$
 (blue dashed lines). }
\label{fig:gp_ep1_w05_pbc}
\end{center}
\end{figure}

\begin{figure}[htbp]
\begin{center}
\includegraphics[width=0.95\linewidth]{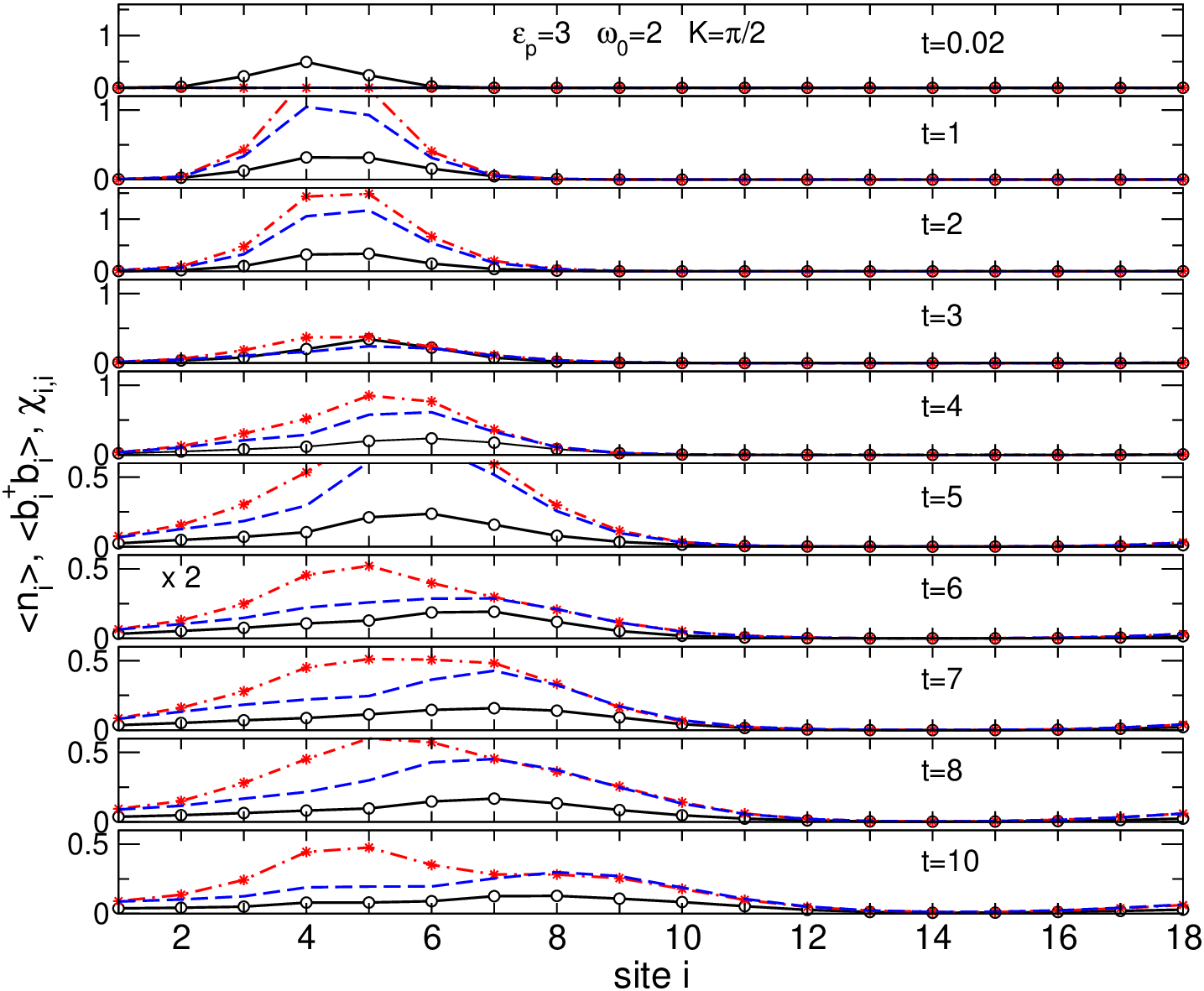}
\includegraphics[width=0.95\linewidth]{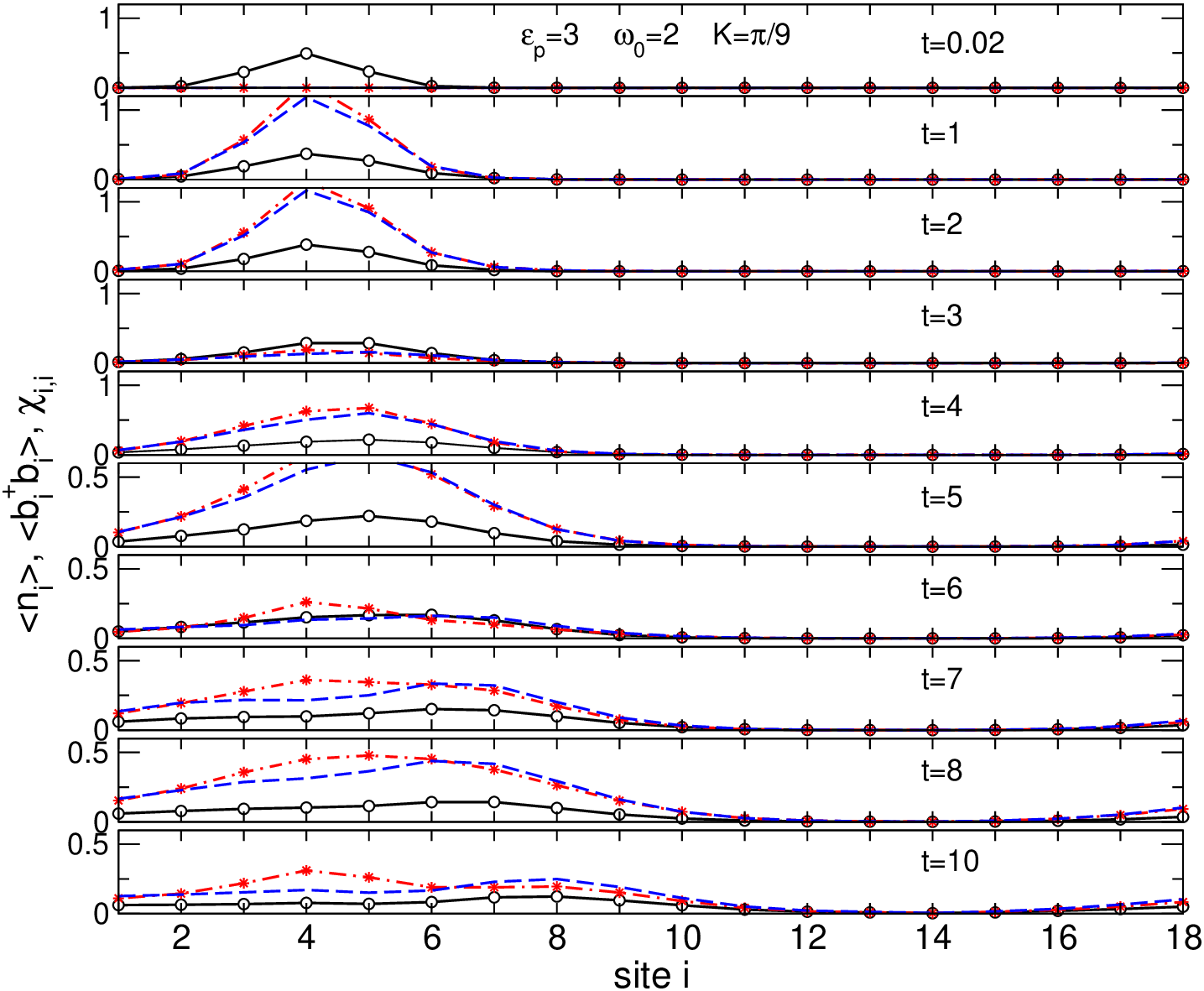}
\caption{(Color online) Spatiotemporal evolution of a free electron Gaussian wave packet 
injected with wave vector $K=\pi/9$ (upper panels)  and $K=\pi/2$ (lower panels) at $t=0$. 
Model parameters are $\varepsilon_p=3$ and $\omega_0=2$. Notations as in Fig.~\ref{fig:gp_ep1_w05_pbc}.}
\label{fig:gp_ep3_w2_pbc}
\end{center}
\end{figure}

Figure~\ref{fig:gp_ep1_w05_pbc} shows snapshots of the local particle densities $\langle n_i\rangle$ 
and phonon numbers  $\langle b_i^\dagger b_i^{}\rangle$ for intermediate EP couplings ($\varepsilon_p=1$ at all sites)
and adiabatic phonon frequencies ($\omega_0=0.5$). 
In addition we included results for  the on-site particle-phonon correlations
\begin{equation}
\chi_{i,i}=\langle n_i^{} (b_i^{} + b_i^+)\rangle\,.
\end{equation}
The particle injected at site 4  is launched to the right
($K=\pi/2$). Shortly after,  the electron is not 
yet dressed and moves nearly as fast as a free particle 
(see black curves in the panels for $t/\tau_e=0.02$, 1, 2, and 3).~\cite{KT07}
At the same time the electron emits (creates) phonons along its path, 
in order to reduce its energy to near the bottom of the band. 
In view of the high initial energy $E_0(t=0)=0$ and an intermediate 
EP coupling strength most of the phonons radiated are uncorrelated 
and therefore continue to stay near the particle's starting point.
Nevertheless the particle drags some phonons with it 
and finally a (coherent) polaron wave packet is formed
characterized by enhanced local particle-phonon correlations  
(see sites 5--7 in the panels for $t/\tau_e=6$--8; note that
the polaronic quasiparticle moves with a reduced velocity~\cite{KT07,WF08}).
Owing to the moderate EP coupling these signatures are rather weak,
however, and are further smeared out when the polaronic 
wave packet dissolves in time.

As Fig.~\ref{fig:gp_ep3_w2_pbc} shows, polaron formation becomes 
more pronounced at larger 
EP couplings ($\varepsilon_p=3$; note the different scale of 
the ordinate compared to 
Fig.~\ref{fig:gp_ep1_w05_pbc}), even if the phonon frequency is enhanced as well ($\omega_0=2$,
non-adiabatic regime). The phonon distribution and enhanced on-site EP correlations indicate that 
more phonons are in the phonon cloud that travels with the particle. Thereby  the polaron inertial mass  
is increased. Again some unbound phonons stay at the point where the particle takes off.
While in the upper graph the particle is injected with an energy of about the phonon energy
above the bottom of the band, its energy is much lower in the lower graph where $K=\pi/9$.
This difference is mainly reflected in the number of unbound phonons; in both cases the polaronic
quasiparticle emerges at about $t/\tau_e=8 \ldots 10$ and shows the same characteristics afterwards.

\begin{figure}[htbp]
\begin{center}
\includegraphics[width=0.95\linewidth]{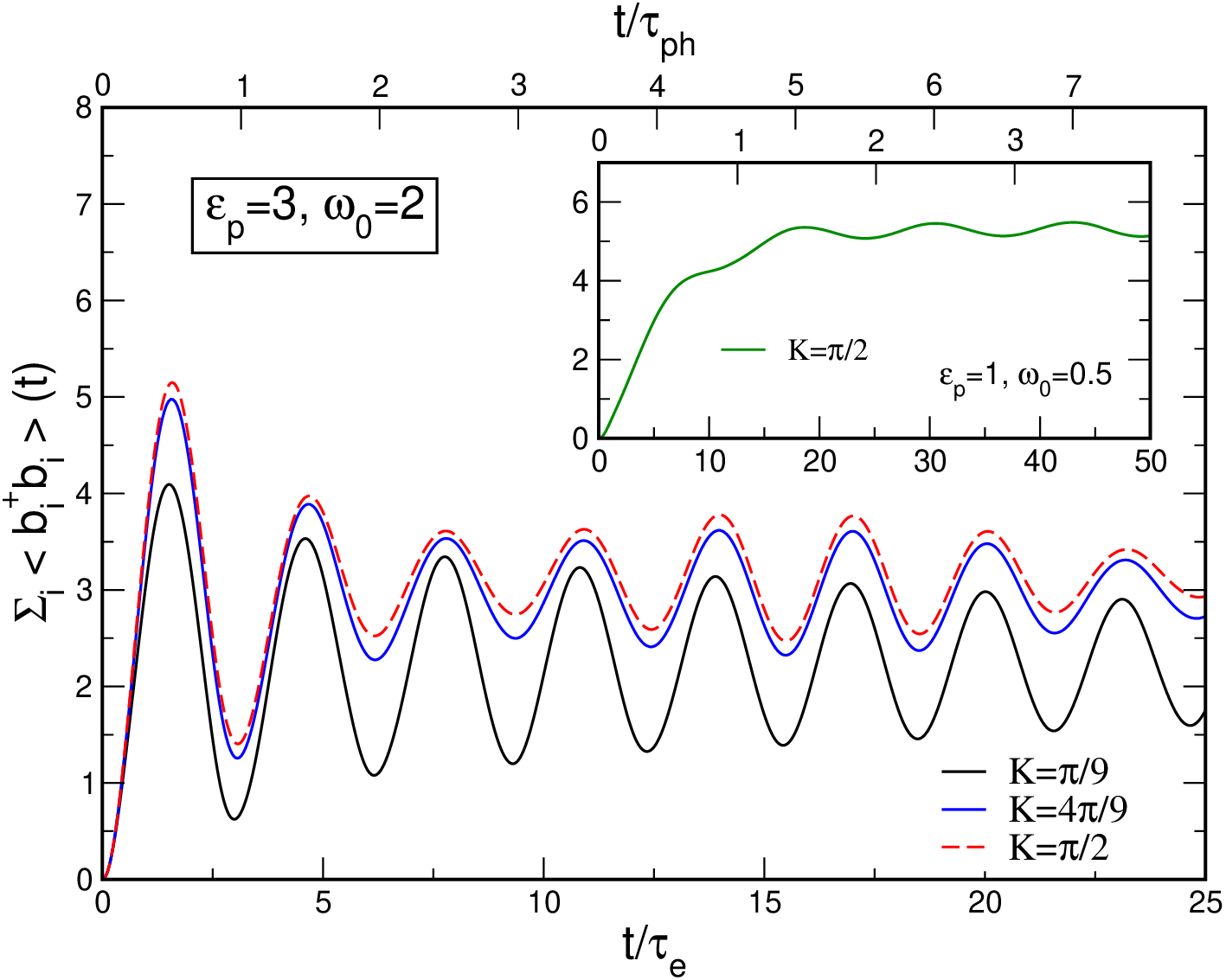}
\includegraphics[width=0.95\linewidth]{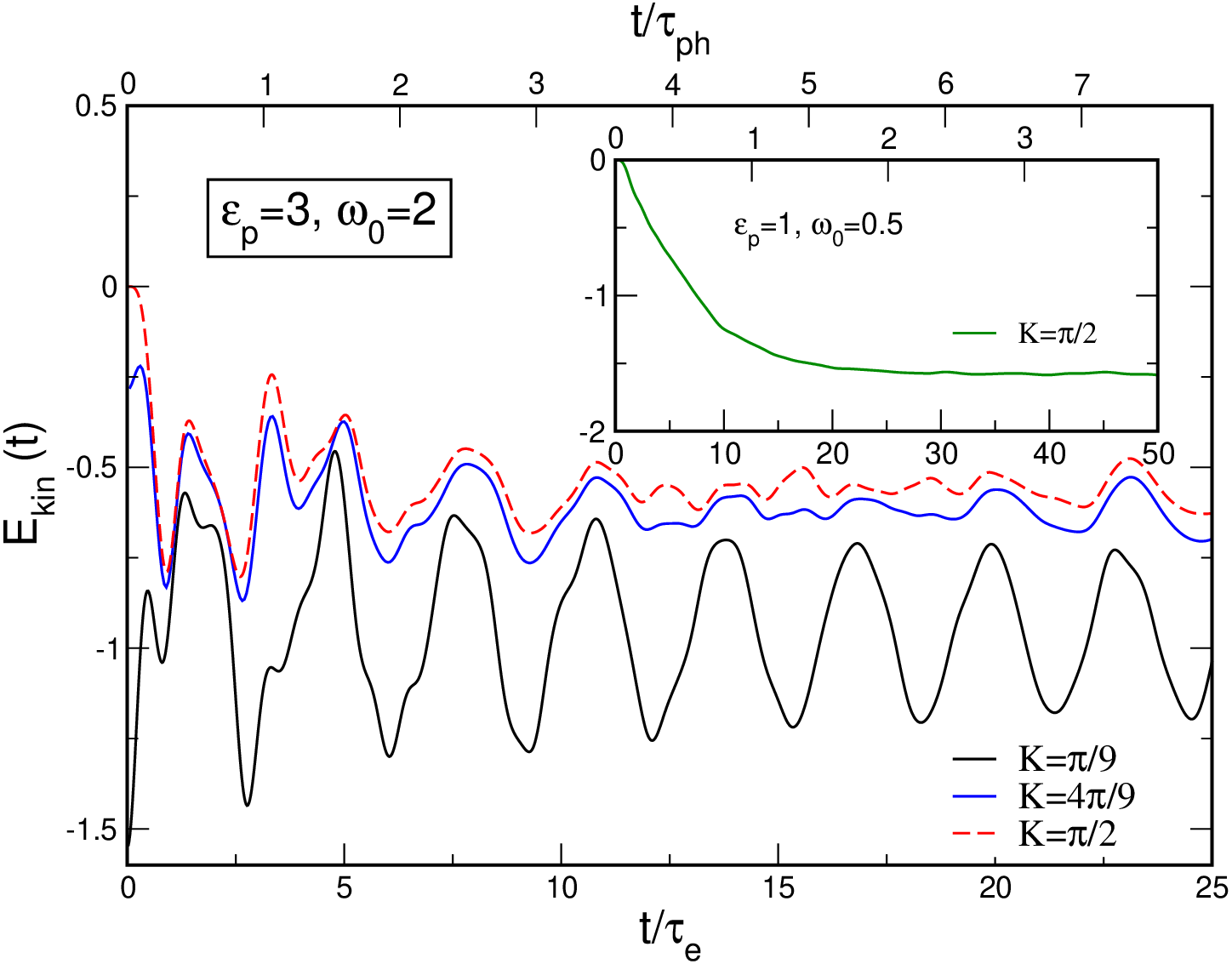}
\caption{(Color online) Upper panel: Time dependence of the total number of phonons excited in the 
molecular crystal chain as the free electron wave packet evolves into a polaron. Lower panel: Time evolution of the 
kinetic energy part  $E_{\rm kin}$.}
\label{fig:ph_tot_Ekin}
\end{center}
\end{figure}

Since the system is not in an eigenstate we expect to find (decaying) oscillations on the time scale 
of $\tau_{ph}$ in the process of polaron formation; at least if $\tau_e$ and $\tau_{ph}$ do not differ too 
much and the EP coupling is not too small. This is illustrated by Fig.~ \ref{fig:ph_tot_Ekin}, 
showing the variation in time of the total number of phonons in the system and of the
kinetic energy part ,
\begin{equation}
E_{\rm kin}= -t_0\sum_i \langle c_i^\dagger c_{i+1}^{}+ {\rm H.c.}\rangle\,.
\end{equation}
For $\omega_0=2$ (main panels) these oscillations can be clearly detected in both quantities. 
The kinetic energy $E_{\rm kin}(\varepsilon_p=3,\omega_0=2)=-1.316$ for the ground state of a polaronic system 
having  the same parameters. 
We find that this value can be much better (periodically) approached, injecting a particle with 
lower energy, i.e. $K=\pi/9$. The minima in $E_{\rm kin}$ are reached when the particle has 
absorbed some phonons. Afterward the particle radiates the phonons again and its kinetic energy increases. 
The oscillations are weaker at $K=\pi/2$. 
While the wave vector of the injected wave packet~(\ref{wave_packet}) is a
continuous variable, finite chains with PBC have only a finite set
of ``allowed'' $K$ vectors. Because  $\pi/2$ is not an allowed 
wave vector of the periodic 18-site system, we included data for $K=4\pi/9$
(located next to $K=\pi/2$) as well, but---as expected---the results 
do not change qualitatively. 
The insets give the total phonon number and kinetic energy
for the parameters of Fig.~\ref{fig:gp_ep1_w05_pbc}, i.e. for the adiabatic case.  
Here we can clearly distinguish two regimes: until $t/\tau_e \sim 6$ 
many unbound phonons were created
to lower the particle's total energy; then polaron formation sets in and the particle attains a kinetic energy
close to the polaron's ground-state kinetic energy $E_{\rm kin}(\varepsilon_p=1,\omega_0=0.5)=-1.823$.

\subsection{Polaron tunneling}
\begin{figure}[htbp]
\begin{center}
\includegraphics[width=0.95\linewidth]{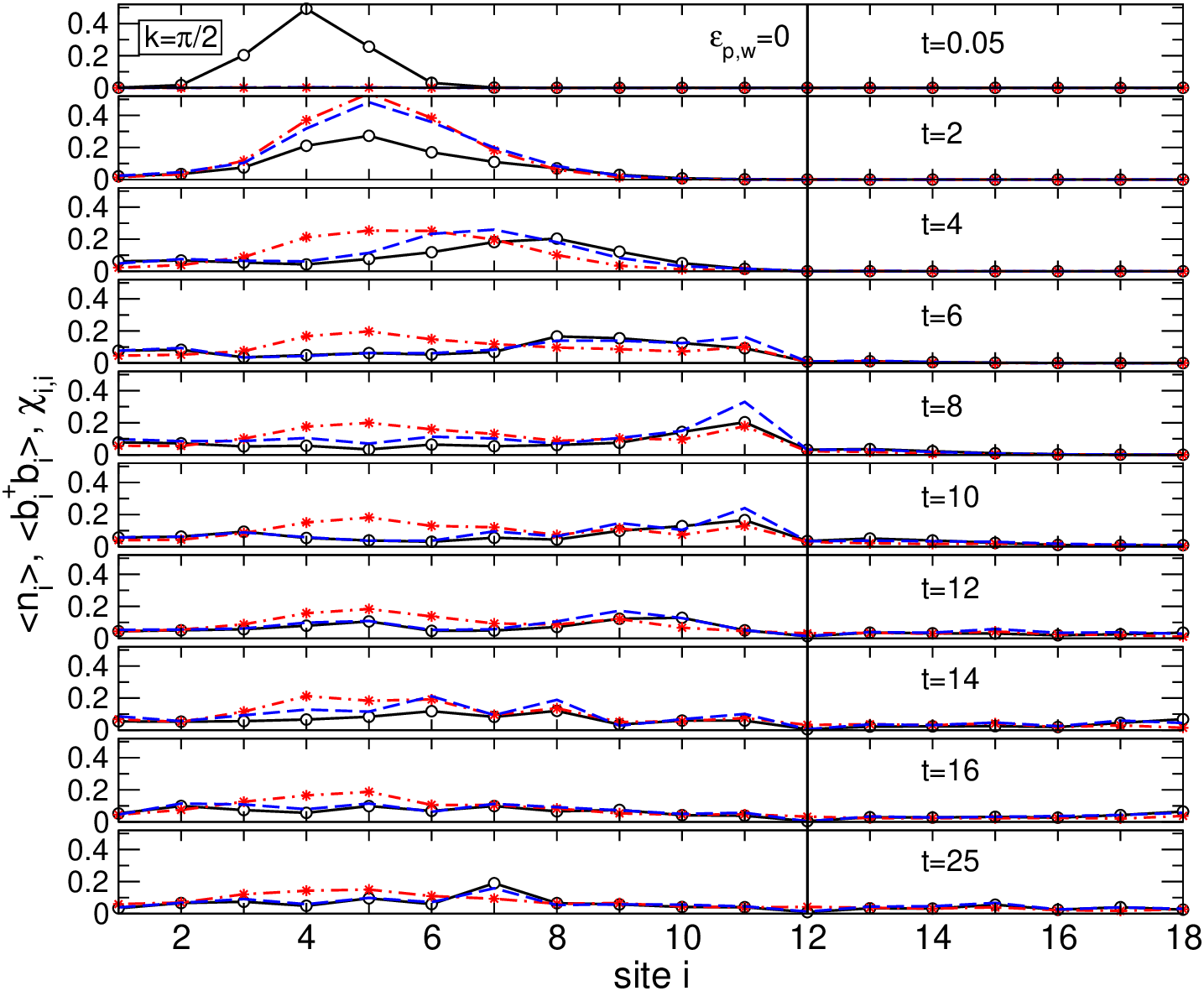}
\includegraphics[width=0.95\linewidth]{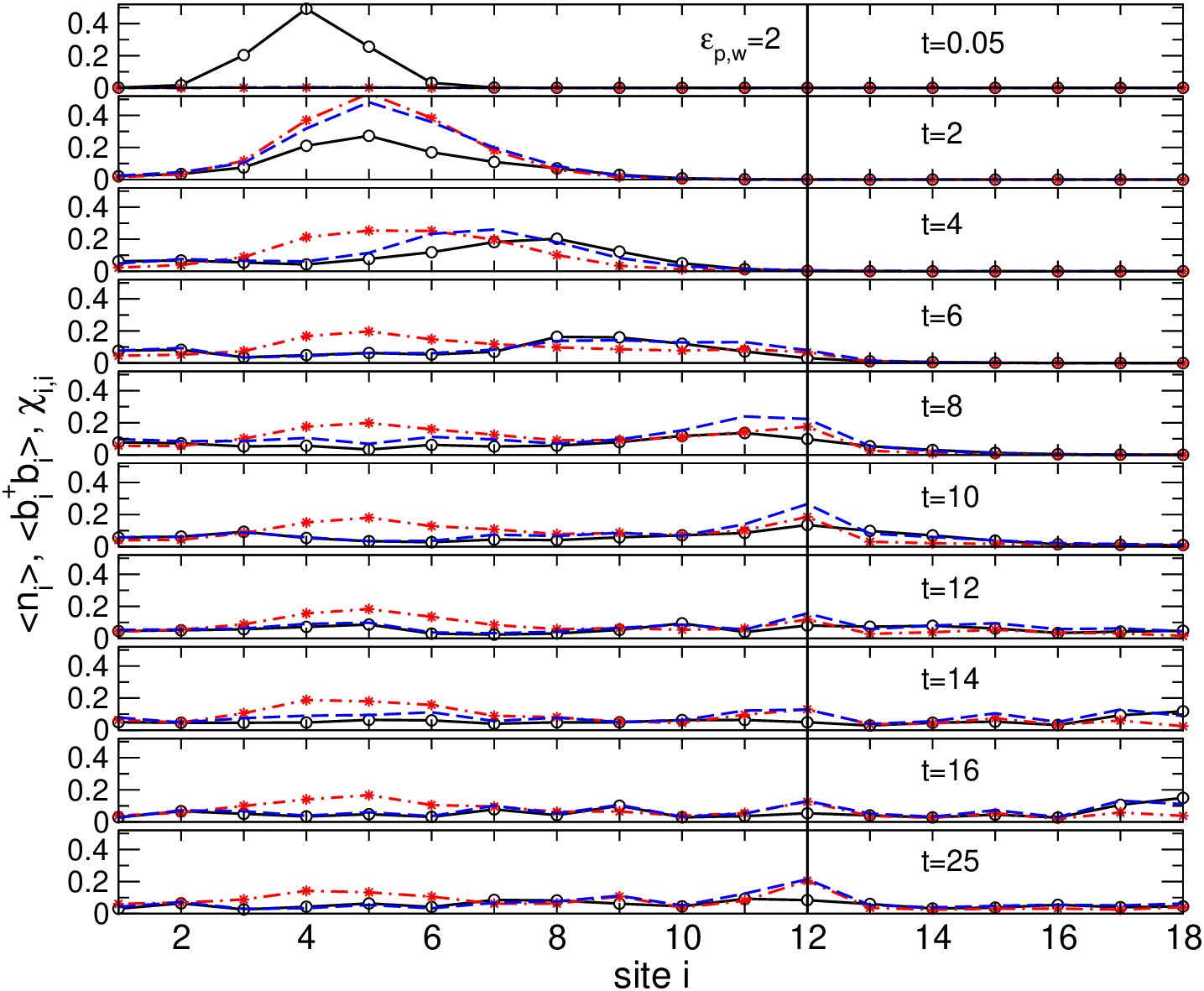}
\includegraphics[width=0.95\linewidth]{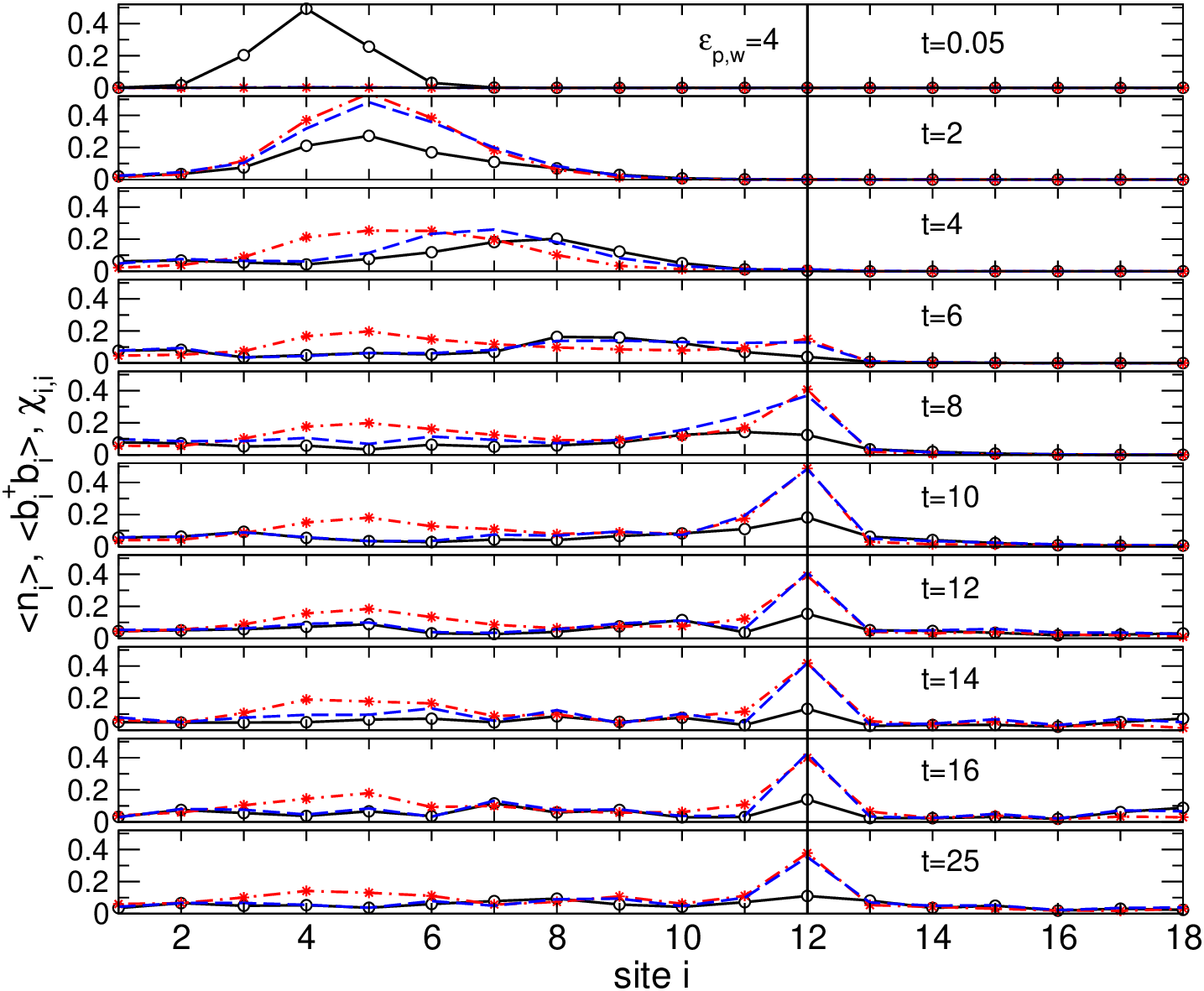}
\caption{(Color online) Quantum dynamics of polaron formation and polaron
tunnelling through a potential barrier $\Delta_{12}=\Delta_w=2$. Model parameters are $\varepsilon_p=1$, $\omega_0=2$.
The barrier or quantum dot is located at site 12 and has a total EP coupling
$(\varepsilon_p +  \varepsilon_{p,w})$. OBCs were used at sites 1 and 18. 
Notation is the same as in Fig.~\ref{fig:gp_ep3_w2_pbc}. 
For further explanation see text.}
\label{fig:gp_ep1_w2_obc}
\end{center}
\end{figure}

\begin{figure}[htbp]
\begin{center}
\includegraphics[width=0.95\linewidth]{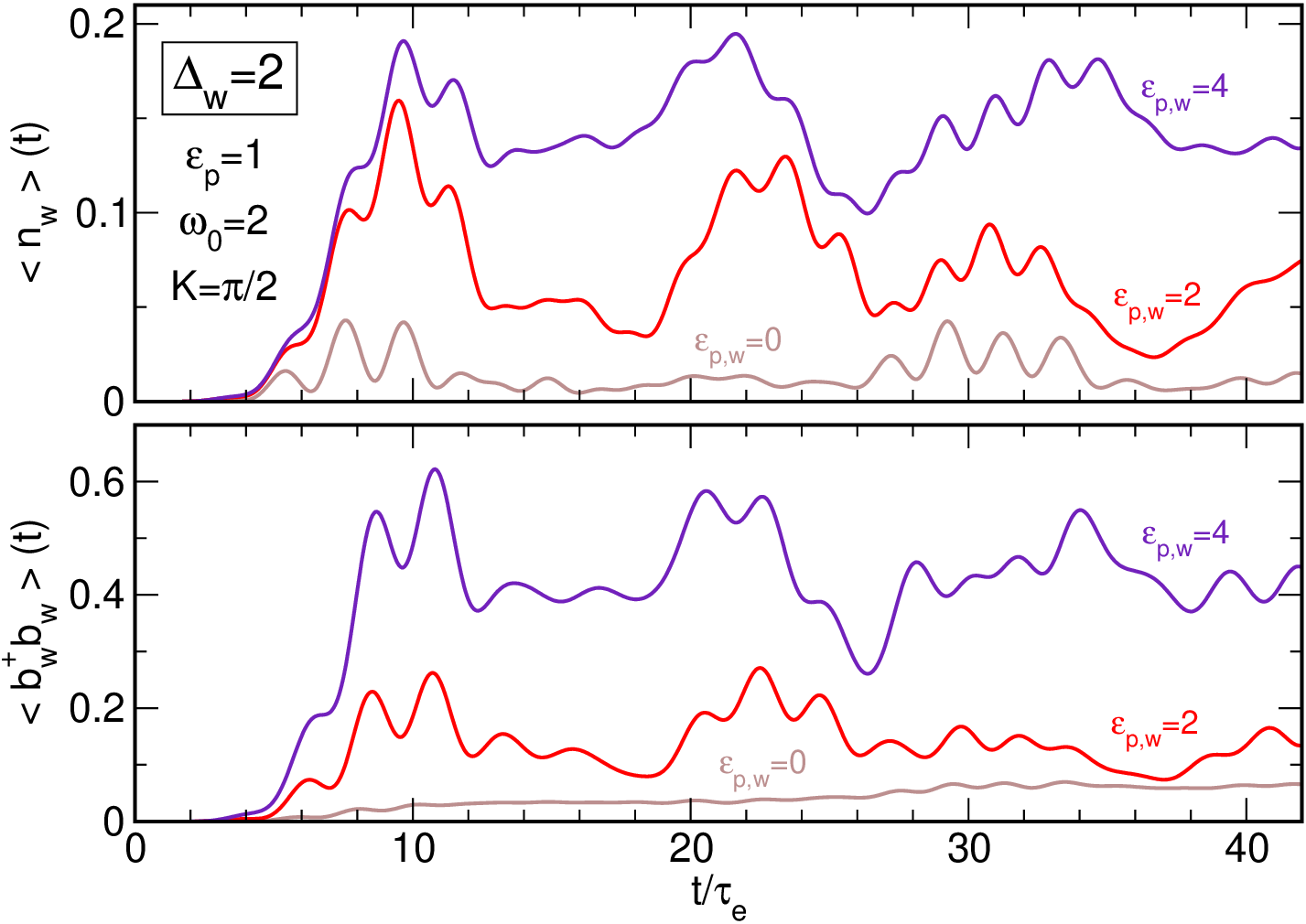}
\caption{(Color online) Time dependence of the particle density (upper panel)
and phonon number (lower panel) at the site of the wall.}
\label{fig:pd_pn_ws}
\end{center}
\end{figure}

\begin{figure}[htbp]
\begin{center}
\includegraphics[width=0.95\linewidth]{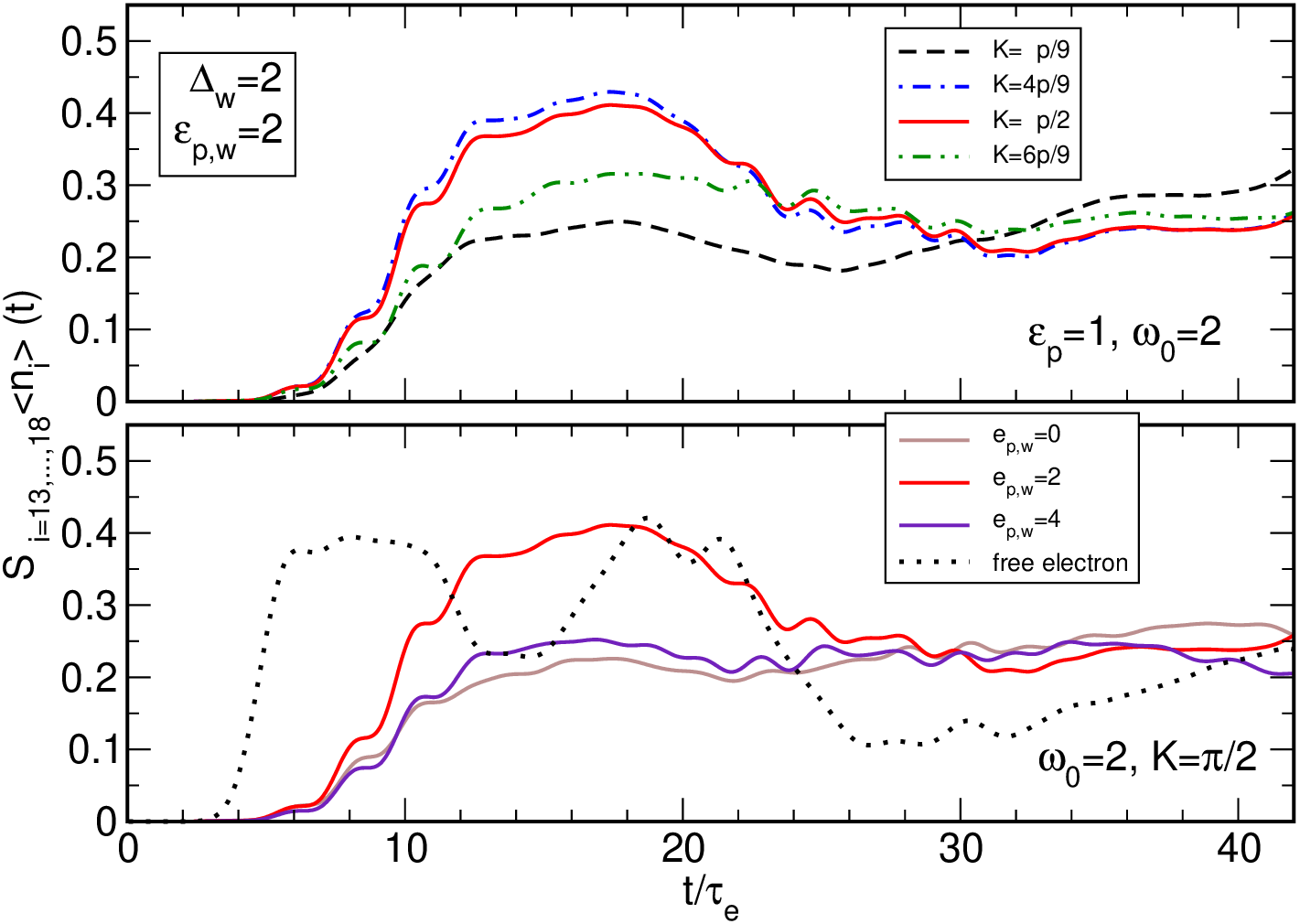}
\caption{(Color online) Cumulated particle density to the right of the potential 
wall at different wave vectors $K$ (upper panel) and various EP couplings $\varepsilon_p$
(lower panel), for the setup of Fig.~\ref{fig:gp_ep1_w2_obc}. }
\label{fig:cpd_rw}
\end{center}
\end{figure}

Finally we investigate the tunneling of a polaronic quasiparticle through 
a potential barrier $\Delta_{12}\equiv\Delta_w=2$ (quantum wall or dot),  
with additional  EP interaction 
$\varepsilon_{p,12}\equiv \varepsilon_{p,w}$.  
For $i\neq 12$ we fix $\Delta_{i}=0$. The other model parameters
are choosen to be $\varepsilon_p=1$ and $\omega_0=2$. 
In the numerics we account for all states with up to $M \le 11$ phonons 
and have checked that in the ground state the weight of basis states containing exactly $M=11$ 
phonons is, for the largest $\varepsilon_{p,w}$, less than $10^{-5}$. 

The wave packet injected has energy $E_0(0)=0$ and moves to the right with
$K=\pi/2$ (see Fig.~\ref{fig:gp_ep1_w2_obc}). 
We apply OBC, so the particle cannot avoid the barrier coming from 
behind. The upper graph describes the situation 
with a barrier at site~12 only. After the polaron is formed at $t/\tau_e 
\sim 6$ it
hits the quantum wall at $t/\tau_e \sim 8-10$ and there is  mostly reflected. Besides this backscattered 
particle current a minor part of the particle tunnels through the barrier, thereby partly stripping 
and recollecting its accompanied phonons (cf. Fig.~\ref{fig:cpd_rw} below). Envisaging a 
vibrating molecular quantum dot located at site~12 (middle panel), an  additive EP interaction $\varepsilon_{p,w}$ 
leads to a local polaronic level shift that softens the barrier. As a result the particle is transmitted
to a much greater extent than in the former case (compare the results for 
$t/\tau_e = 14-16$).
If the quantum dot possess a very strong EP interaction the polaron digs  at 
the dot site and stays there for a long time (see lower panel). Then of course both the reflected
and transmitted particle current is low. 

Figure~\ref{fig:pd_pn_ws} gives the temporal  variation of particle density and phonon number
at the quantum wall/dot site. Obviously the phonons somewhat lag behind the electron (retardation
effect). During the tunneling process the phonon number strongly fluctuates. The $\varepsilon_{p,w}=4$ curve
clearly signals the self-trapping of the electron at the dot site. The second bump-series is due to 
electronic contributions retaining after being reflected at the system's boundary (site~18).  

The total transmitted electron density is displayed in Fig.~\ref{fig:cpd_rw}
for $\varepsilon_{p,w}=\Delta_w$ and various momenta of the injected wave packet 
(upper panel) as well as for different $\varepsilon_{p,w}$ at $K=\pi/2$ (lower panel).
Of course a higher initial energy enhances the transmission through
the tunnel barrier (compare the $K=\pi/2$ and $K=\pi/9$ curves).
The lower panel  first of all shows the time delay of the polaron
in reaching the barrier compared to a free particle (dotted line). 
More notably, we observe that for  $\varepsilon_{p,w}=2$
the transmission is as high as for free particles, despite the fact
that the particle is dressed by phonons in a significant way.  Note that the dimensionless 
EP coupling parameters at the dot site are $3/2 t_0$ and $g_{12}^2=1.5$. 
This points toward the importance of vibration-mediated tunneling processes
(doorway vibrons).~\cite{FWLB08}

\section{Summary}
In this work, we have presented an efficient numerical method to calculate the time 
evolution of the many-body wave function of an interacting electron-phonon system. 
The approach is based on Chebyshev moment expansion, applied to the time evolution 
operator.  We focused on the process of small polaron formation in finite low-dimensional 
quantum structures described by a generalized Holstein Hamiltonian. Both electron and 
phonon quantum dynamics were treated exactly.

We first started from a non-interacting ground state and analyzed the real-time dynamics of  
the particle density and phonon number after a sudden switching-on of the electron-phonon 
coupling at a single oscillatory (molecular quantum dot)  site. As a consequence of this 
interaction quench the originally free particle can be trapped at the ``impurity'' site after a while.
The  self-trapping process differs in nature for the adiabatic and anti-adiabatic
regimes of small and large phonon frequencies, respectively. In the former case, where
the phonons are slow and retardation effects play an important role, a static lattice distortion
evolves that causes an effective attractive potential for the electron. As a result a 
Holstein polaron is formed. In the latter case phonons can follow the electron motion almost 
instantaneously. Hence we observe very fast phonon emission and re-absorption processes, which---at large EP interaction strengths---give rise to a dynamical dressing of the charge carrier
that enhances the particle's mass and finally leads to its immobilization. In both cases
the phonon distribution function signals the existence of excited bound polaron-phonon
states. Since our initial state is not an eigenstate of the interacting system,  
we observe the phenomenon of recurrence at later times.  

Next we launched a free-electron Gaussian wave packet in a one-dimensional system, 
subjected to EP coupling at every site. The injected bare particle is found to
radiate phonons to lower its energy to near the bottom of the band. Thereforre part of
the phonons stay near the electron's starting point and---if the EP coupling is sufficiently
strong---another part of the phonons will be embedded in a phonon cloud attached to the (moving) 
particle. The latter polaron quasiparticle formation process takes a period of time 
that depends on the characteristic electron and phonon times scales, the EP interaction 
strength, and the initial conditions in a very sensitive way.  We agree with the findings 
of previous work~\cite{KT07} that the question of how long it takes a polaron to form, 
has no simple answer, because there are multiple time scales in the dynamics.  

In the last part we investigated the transmission of a polaron through a quantum wall
or vibrating quantum dot. Depending  on the barrier height to electron-phonon interaction 
strength ratio, and the characteristic electron and phonon times scales, we found
opposed behaviors: strong reflection; phonon-mediated tunneling; and intrinsic localization of the polaron. 
Most notably we showed that if the polaronic level lowering just compensates the
repulsive dot potential and the electron and phonon time scales are comparable,
a rather heavy small polaron, regardless of its  phonon cloud, tunnels like a free electron, On the other hand, if there is a mismatch between both quantities ,
we observe strong phonon fluctuations at the dot site and transport through the quantum 
dot becomes significantly suppressed. This might motivate further investigations
of deformable quantum dot systems, e.g. with respect to applications as a current switch.

In conclusion, we have demonstrated that polaron formation is a subtle non-linear dynamical process
which is affected by multiple time/energy scales. The proposed long-time
Chebyshev expansion method---in combination 
with exact diagonalization techniques---is capable of  addressing 
such complex problems, which raises the expectation  that our approach 
can also be used to study the time-evolution of quasiparticles in more general situations. 
 
\acknowledgements
The authors would like to thank A. Alvermann,  J.~Loos, G. Schubert, and S. A. Trugman
for valuable discussions. HF and GW acknowledge the hospitality at 
Los Alamos National Laboratory. This work was supported by KONWIHR Bavaria (HF, GW) 
and the US Department of Energy (ARB). Numerical calculations were performed at the 
LRZ Munich.

\end{document}